\documentclass[epj]{svjour}
\usepackage{graphicx}
\usepackage{amsmath}
\usepackage{amssymb}
\usepackage{slashed}
\usepackage{hyperref}
\usepackage{float}
\RequirePackage{fix-cm}
\renewcommand{\figurename}{Figure}
\begin{document}
\title{Vacuum Polarization Effects from Dark Fermion Loops in $Z'$ Mediator Models and their Impact on Angular Observables in $B \to K^* \mu^+\mu^-$ Decays}
%\subtitle{Do you have a subtitle?\\ If so, write it here}
\author{Suhani Gupta\inst{1} and Sunil Dogra\inst{2}}% etc
%\thanks is optional - remove next line if not needed
%\thanks{\emph{Present address:} Insert the address here if needed}%
                     % Do not remove
%
\offprints{S. Gupta (suhanigupta0906@gmail.com), S. Dogra (sdogra@cern.ch)} % Insert a name or remove this line
\institute{Amity International School, Sector 46, Gurugram, 122002 Haryana, India\and Centre for High Energy Physics, Kyungpook National University 80, Daehak-ro, Buk-gu, 41566 Daegu, South Korea}
\date{Received: \today / Revised version: \today}
% The correct dates will be entered by Springer
%
\abstract{ 
Rare flavour changing neutral current decays such as $B \to K^{*}\mu^{+}\mu^{-}$ provide sensitive probes of physics beyond the Standard Model (SM), being forbidden at tree level and arising only through loop induced processes. Angular analyses yield observables sensitive to short distance dynamics. The optimized observable $P'_5$ shows a persistent $\sim 3\sigma$ deviation from SM predictions in $q^{2}\in[1.1,6.0]\,\mathrm{GeV}^{2}$, motivating modified Wilson coefficients. A theoretically motivated framework is investigated in which vacuum polarization of a Dirac dark fermion $\chi$ modifies the propagator of a heavy $Z'$ boson, inducing a momentum dependent contribution to $C_{9}(q^{2})$. The loop corrected propagator is incorporated into the effective weak Hamiltonian for $b\to s\ell^{+}\ell^{-}$. Bin averaged angular observables are computed with \texttt{flavio}, and $34{,}997$ parameter points are scanned over different chiral structures of the $Z'$ couplings. A $\chi^{2}$ analysis identifies preferred regions that are vector dominated and chirally asymmetric. The best fit solution significantly alleviates the $P'_5$ tension, reducing the deviation in the central $q^{2}$ bin to $0.12\sigma$, an improvement of approximately $96\%$ relative to the SM. This improvement is driven by tree level $Z'$ exchange, which induces $\Delta C_{9}\simeq-1.2$, consistent with global fits and compatible with other angular observables and $B_{s}\to\mu^{+}\mu^{-}$. Dark fermion vacuum polarization adds a subleading $q^{2}$ dependence to $C_{9}$; it does not dominate the fit, but supplies a distinctive threshold structure testable with finer binning. These results show that a $Z'$ portal to a dark fermion provides a phenomenologically consistent description of the $P'_5$ anomaly and a concrete benchmark for LHCb Run~3 analyses.
\PACS{ Decays of bottom mesons, Dark matter, Neutral currents and Neutral currents}
% end of PACS codes
}%end of abstract
\maketitle
\section{Introduction}
\label{intro}

The study of \(B\) mesons occupies a central place in flavour physics. These mesons, containing a bottom quark, provide access to flavour changing neutral current (FCNC) processes, which are forbidden at tree level in the Standard Model (SM) and arise only through higher order loop contributions. This suppression makes them highly sensitive probes of possible new physics (NP) effects.  In recent years, rare \(B\) meson decays \cite{PhysRevLett.109.101802,PhysRevD.88.072012,LHCb:2015gmp,PhysRevD.92.072014,PhysRevD.94.072007,PhysRevLett.118.211801,PhysRevD.97.012004,PhysRevLett.120.171802,PhysRevD.97.072013,PhysRevD.108.032002,Aaij_2023} have become a key focus in the search for NP. Among the most prominent observables are the lepton flavour universality (LFU) ratios $R_K$ and $R_{K^*}$ in the decays $B \to K \ell^+ \ell^-$ and $B \to K^* \ell^+ \ell^-$, respectively. In the SM, these ratios are expected to be close to unity due to LFU. Earlier measurements by the LHCb collaboration indicated deviations below unity~\cite{LHCb:2019hip}, generating significant interest as potential signs of NP. However, with the full Run~2 dataset, updated analyses reported in ~\cite{PhysRevD.108.032002,Aaij_2023} found these ratios to be consistent with SM expectations.

Despite this agreement, tensions persist in certain angular observables of $b \to s \ell^+ \ell^-$ transitions ~\cite{LHCb:2021zwz}. In particular, the observable $P'_5$, extracted from the full angular analysis of $B \to K^* \mu^+ \mu^-$ decays, remains one of the most sensitive probes of short distance physics encoded in the Wilson coefficients $C_9$ and $C_{10}$ ~\cite{LHCb:2025mqb}. Constructed from ratios of transversity amplitudes, $P'_5$ provides a particularly clean and robust probe of NP effects in FCNC processes \cite{lhcb2025angular}. Global fits to map theoretical values of $P'_5$ with collider data have been consistently interpreted as a possible negative shift in $C_9$ ~\cite{crivellin2023diquark,Altmannshofer:2021C9fit,Capdevila:2017bsm}. The latest results from the LHCb collaboration report deviations in observables such as $F_L$, $P'_5$, $S_5$, and $S_9$ at the level of $1.5$--$2\sigma$~\cite{lhcb2025angular}, with a local deviation in $P'_5$ approaching $\sim 3\sigma$ in the theoretically clean region $q^2 \in [1.1,\,6.0]~\mathrm{GeV}^2$. 

\begin{figure}[t]
\centering
 \includegraphics[width=0.7\linewidth]{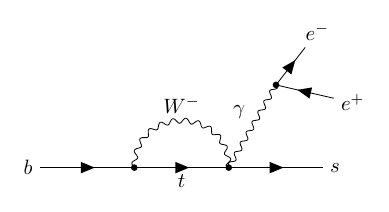}
\caption{Penguin diagram for the \(b \to s \gamma^* \to e^+ e^-\) transition via a virtual \(W\) boson and top quark loop.}
\label{fig:bsgamma-penguin}
\end{figure}

Current theoretical efforts aim to explain these deviations within consistent extensions of the SM. Broadly, two classes of scenarios are considered, loop level contributions from dark sector particles and mediator induced effects such as those arising from heavy $Z'$ gauge bosons or leptoquarks \cite{herrero2018dark,di2017minimal,chiang2016z,bhatia2022frugal}. Simple $Z'$ models require sizeable flavour changing $b$--$s$ couplings to account for the observed shift in $C_9$, but such couplings are tightly constrained by $B_s$--$\bar{B}_s$ mixing and high $p_T$ dilepton searches at the LHC. On the other hand, purely loop induced explanations involving dark sector particles are typically suppressed and often require large couplings or light states that are disfavoured by experimental bounds. These challenges motivate more general scenarios in which the required contributions can be generated while remaining consistent with existing flavour, collider and dark sector constraints.

Additionally, when dark sector particles are considered, most current dark matter (DM) detection experiments search for candidates known as Weakly Interacting Massive Particles (WIMPs). These candidates are typically assumed to have masses in the GeV-TeV range~\cite{liu2026search} and are already strongly constrained. Collider and direct detection limits depend on the WIMP interactions with quarks, such constraints can therefore be alleviated in scenarios where the dark sector communicates with the SM through a mediator rather than through direct interactions~\cite{wang2025prospects}.

A well motivated framework that connects the SM with a hidden sector is an additional Abelian gauge symmetry \(U(1)'\). In this framework both SM fermions and dark sector particles are charged under the new symmetry, and the corresponding gauge boson \(Z'\) mediates interactions via renormalizable current--current couplings. After symmetry breaking the \(Z'\) acquires a mass. Integrating it out at energies below its mass scale generates effective four fermion operators relevant for \(b\to s\ell^+\ell^-\) transitions. 

Therefore the \(Z'\) framework has been employed in two distinct sectors, flavour physics and dark-sector physics. Because \(Z'\) bosons appear in both dark-sector portal models and explanations of \(B\) meson decays, a natural question arises as to whether a dark fermion can affect \(B\) meson decay transitions. Most existing analyses assume that \(Z'\) contributions arise purely at tree level. 

In a complete quantum field theoretical framework, however, this approximation is incomplete, as radiative corrections are inevitably generated. In particular, vacuum polarization diagrams involving virtual dark fermions contribute to the \(Z'\) self energy, thereby modifying the propagator. Consequently, the \(Z'\) propagator acquires a non trivial momentum dependence, leading to a \(q^2\) dependent modification of the effective Wilson coefficient \(C_9\). Such momentum dependent effects can be directly probed through angular observables in rare \(B\) meson decays, particularly the observable \(P'_5\)~\cite{Peskin:1995ev}.

In the present work, a scenario is examined in which the \(Z'\) boson is dressed by vacuum polarization effects arising from a dark fermion loop. This setup constitutes an intermediate framework between simple tree level mediator models and purely loop induced dark sector scenarios. The resulting momentum dependent correction can generate a localized shift in \(C_9\) without requiring large flavour violating couplings, thereby avoiding stringent constraints from flavour and collider measurements.
%Above is correct
\section{Theoretical Framework} 
\label{theoryFramework}

This study develops a novel dark sector framework for the $Z'$ mediator to explain the persistent deviations observed in the angular observables of $B \to K^* \mu^+ \mu^-$ decays. This approach incorporates quantum loop effects from dark Dirac fermions directly into the flavour changing amplitude within an effective field theory description of FCNC transitions. In the SM, the $B \to K^* \mu^+ \mu^-$ decay proceeds only at the one loop level through electroweak penguin diagrams (Fig.~\ref{fig:bsgamma-penguin}). This loop suppression makes the process highly sensitive to NP contributions.

The proposed NP model is LFU violating by coupling preferentially to muons and neutrinos while leaving electrons and taus unaffected. It consists of three main components. A new neutral vector boson $Z'_\mu$ arises from an additional $U(1)'$ gauge symmetry. This $Z'$ mediator couples to both SM fermions and dark sector fields, thereby modifying the Wilson coefficients $C_9$ and $C_{10}$ that govern $b \to s \ell^+ \ell^-$ transitions, and possesses both left- and right-handed couplings. A Dirac fermion $\chi$ serves as the DM candidate. Although $\chi$ does not couple directly to SM quarks or leptons, virtual $\chi\bar{\chi}$ pairs induce vacuum polarization corrections to the $Z'$ propagator, which in turn affect low energy flavour observables. The SM sector provides the baseline description of $b \to s \ell^+ \ell^-$ transitions. While vacuum polarization effects from SM fermions are already included in SM predictions, the additional dark sector fermion offers a well motivated extension capable of simultaneously addressing the \(B\) meson anomalies and DM phenomenology.

\begin{figure}[t]
\centering
\includegraphics[width=0.7\linewidth]{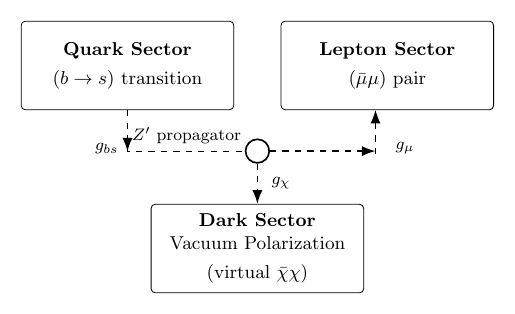}
  \caption{Schematic representation of the effective couplings: the \(Z'\) mediates the \(b\to s\) transition (quark sector) and couples to the muon pair (lepton sector), while a dark fermion loop (\(\chi\bar{\chi}\)) contributes via vacuum polarization.}
\label{fig:zp-schematic}
\end{figure}

\begin{figure}[thbp]
\centering
\includegraphics[width=0.7\linewidth]{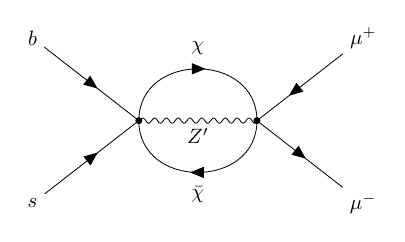}

\caption{Feynman diagram for the \(b \to s \mu^+ \mu^-\) transition mediated by a \(Z'\) boson, with a dark fermion loop (\(\chi \bar{\chi}\)) contributing to the vacuum polarization.}
\label{fig:zp-dark-loop}
\end{figure}

The dark Dirac fermion $\chi$ couples vectorially to the $Z'$ through a conserved current. This coupling ensures the absence of gauge anomalies and allows the $Z'$ to act as a portal between the visible and dark sectors, generating loop level modifications to the gauge boson self energy. The resulting portal structure is shown schematically in Fig.~\ref{fig:zp-schematic}: the $Z'$ couples the $b\to s$ current to the muon pair while a virtual $\chi\bar\chi$ loop corrects the propagator. The corresponding Feynman diagram, with the dark fermion vacuum polarization insertion on the $Z'$ line, is shown in Fig.~\ref{fig:zp-dark-loop}.

The full Lagrangian extends the SM by the additional $U(1)'$ gauge symmetry as:
\begin{equation}
\mathcal{L} = \mathcal{L}_{\rm SM} + \mathcal{L}_{Z'} + \mathcal{L}_\chi + \mathcal{L}_{\rm int}.
\end{equation}
The kinetic and mass terms for the new gauge boson are given by:
\begin{equation}
\mathcal{L}_{Z'} = -\frac{1}{4} Z'_{\mu\nu} Z'^{\mu\nu} + \frac{1}{2} M_{Z'}^2 Z'_\mu Z'^\mu.
\end{equation}
where $Z'_{\mu\nu} = \partial_\mu Z'_\nu - \partial_\nu Z'_\mu$, and $m_{Z'}$ is the mass of the new boson, typically arising from the spontaneous breaking of the $U(1)'$ symmetry in a hidden sector.

 The dark fermion $\chi$ is modeled as a Dirac field carrying a $U(1)'$ charge and its Lagrangian includes a
minimal coupling to the $Z'$  through a gauge interaction:
\begin{equation}
\mathcal{L}_\chi = \bar{\chi} (i \gamma^\mu \partial_\mu - m_\chi) \chi + g_\chi \bar{\chi} \gamma^\mu \chi \, Z'_\mu,
\end{equation}
where $m_\chi$ denotes the dark fermion mass and $g_\chi$ its coupling strength to the $Z'$. This interaction term automatically arises from the covariant derivative formalism, $D_\mu = \partial_\mu - i g_\chi Z'_\mu$, and is responsible for generating vacuum polarization corrections to the $Z'$ propagator through quantum loops of $\chi$. The interaction terms connecting the NP to the SM fields take the form:

\begin{equation}
\begin{aligned}
\mathcal{L}_{\rm int} = Z'_\mu \Big[ 
& \bar{s} \gamma^\mu (g_{bs}^L P_L + g_{bs}^R P_R) b \\
& + \bar{\mu} \gamma^\mu (g_\mu^L P_L + g_\mu^R P_R) \mu \\
& + g_\chi \bar{\chi} \gamma^\mu \chi \Big].
\end{aligned}
\end{equation}

where $g_{bs}^{L}$ and $g_{bs}^{R}$ represent the left- and right-handed quark couplings to the $Z'$ mediator, while $\chi$ couples to $Z'$ through a purely vector interaction. The NP Lagrangian is given as:

\begin{equation}
\begin{aligned}
\mathcal{L}_{\text{NP}} =\;
& -\frac{1}{4} Z'_{\mu\nu} Z'^{\mu\nu}
+ \frac{1}{2} M_{Z'}^{2} Z'_\mu Z'^\mu \\
& + \bar{\chi}\left(i\gamma^\mu \partial_\mu - m_\chi\right)\chi \\
& + Z'_\mu \Big[
\bar{s}\gamma^\mu (g_{bs}^L P_L + g_{bs}^R P_R)b \\
&  + \bar{\mu}\gamma^\mu (g_\mu^L P_L + g_\mu^R P_R)\mu \\
&  + g_\chi\, \bar{\chi}\gamma^\mu \chi
\Big].
\end{aligned}
\end{equation}

At low energies the dynamics are described by the effective Hamiltonian:
\begin{equation}
\mathcal{H}_{\rm eff} = -\frac{4 G_F}{\sqrt{2}} V_{tb} V_{ts}^* 
\sum_i \Big( C_i(\mu) \mathcal{O}_i(\mu) + C_i'(\mu) \mathcal{O}_i'(\mu) \Big).
\end{equation}
Here $C_i(\mu)$ are the Wilson coefficients and $\mathcal{O}_i(\mu)$ are the local four fermion operators. In the presence of NP with right handed quark currents, the effective Hamiltonian must be extended beyond the SM operator basis to include chirality flipped operators. Such operators arise naturally in models with non universal or asymmetric couplings of a heavy $Z'$ mediator and are essential for consistently describing left right asymmetric scenarios. For the interaction considered in this work, the relevant operators for this analysis are:

\begin{align}
\mathcal{O}_9 &= (\bar{s} \gamma_\mu P_L b)(\bar{\mu} \gamma^\mu \mu), \quad \\
\mathcal{O}_9' &= (\bar{s} \gamma_\mu P_R b)(\bar{\mu} \gamma^\mu \mu), \\
\mathcal{O}_{10} &= (\bar{s} \gamma_\mu P_L b)(\bar{\mu} \gamma^\mu \gamma_5 \mu), \quad \\
\mathcal{O}_{10}' &= (\bar{s} \gamma_\mu P_R b)(\bar{\mu} \gamma^\mu \gamma_5 \mu).
\end{align}

Where $P_{L,R} = (1 \mp \gamma_5)/2$. These operators govern the angular and kinematic structure of the decay. The primed operators correspond to right handed quark currents and do not appear in the SM, but contribute in generic $Z'$ extensions
 with chiral asymmetry. The resulting effective contributions to the short distance physics are encoded in the Wilson coefficients $C_9$, $C'_9$, $C_{10}$, $C'_{10}$. Different chiral realizations of the $Z'$ mediator correspond to distinct patterns of these coefficients, which are systematically explored in the following sections. These operators and their associated Wilson coefficients receive NP contributions that modify the angular distribution of the decay, particularly the observable $P_5'$. Once the $Z'$ mediator couples to both SM fermions and the dark fermion $\chi$, quantum corrections can alter how it propagates between interaction points. At tree level, the $Z'$ propagator has the following structure ($m_{z'}$ is the mass of the mediator $\eta_{\mu ,\nu}$ is the Minkowski metric):

\begin{equation}
D^{(0)}_{\mu\nu}(q) = \frac{-i \eta_{\mu\nu}}{q^2 - M_{Z'}^2 + i\epsilon}.
\end{equation}

When virtual DM pairs $\chi \bar\chi$ can temporarily form in vacuum, they modify the propagation by generating a vacuum polarization term $\Pi(q^2)$. The corrected propagator becomes:
\begin{equation}
D_{\mu\nu}(q) = \frac{-i \eta_{\mu\nu}}{q^2 - M_{Z'}^2 + \Pi(q^2) + i\epsilon}.
\end{equation}

To isolate the physically observable effects, only the transverse part of the propagator is considered, since gauge invariance ensures that only this component contributes to on shell amplitudes:
\begin{equation} \label{effprop}
    D_{\mu \nu}(q)\simeq \frac{-i}{q^2-m_{Z'}^2+\Pi(q^2)}(\eta_{\mu \nu} - \frac{q_\mu q_\nu}{q^2})
\end{equation}

To account for the corrections induced by the dark fermion $\chi$ to the $Z'$ mediator propagator, the one-loop vacuum polarization tensor $\Pi^{\mu\nu}(q)$ was evaluated. This tensor describes the modification of the $Z'$ propagator arising from virtual $\chi\bar{\chi}$ fluctuations. At one loop, the vacuum polarization tensor is given by:
{\footnotesize 
\begin{equation}
\begin{aligned}
i \Pi^{\mu\nu}(q) = -(-i g_\chi)^2 \int \frac{d^4k}{(2\pi)^4} \frac{\operatorname{Tr}\bigl[ \gamma^\mu (\slashed{k} + m_\chi) \gamma^\nu (\slashed{k} + \slashed{q} + m_\chi) \bigr]}{(k^2 - m_\chi^2) \bigl[ (k+q)^2 - m_\chi^2 \bigr]}
\end{aligned}
\end{equation}
}

Here k is the loop momentum, q is the total momentum carried by the propagator. This tensor is transverse due to gauge invariance, thus $q_\mu \Pi^{\mu \nu}(q)=0$ which allows the above integral to ultimately decompose as:

\begin{equation}
    \Pi^{\mu \nu}(q)=(q^2g^{\mu \nu}-q^\mu q^\nu)\Pi(q^2)
\end{equation}

The self energy function $\Pi(q^2)$ is a scalar function encoding the entire physical effect of the loop.
\begin{equation}
    \Pi_{\mu \nu}(q^2)= (g_{\mu \nu}- \frac{q_{\mu}q_{\nu}}{q^2})
\end{equation}

To evaluate $\Pi(q^2)$, the Feynman Parameterization Trick is used to combine the two denominators into a single term. Once the momentum integration is carried out using dimensional regularization, the intermediate form can be expressed as:
\begin{equation}
\begin{aligned}
    \Pi(q^2)=&-\frac{g_{\chi}^2}{2\pi^2}\int^1_0 dx\text{ } x(1-x) [\frac{2}{\epsilon}+ \gamma_E+ \\ & \ln(4\pi) 
    -\ln \frac{\Delta(x,q^2)}{\mu^2}+ constant]
    \end{aligned}
\end{equation}

Here, $\mu$ denotes the renormalization scale, and $\Delta(x,q^2)=m_{\chi}^2-x(1-x)q^2$. On shell renormalization is employed, with the subtraction performed at $q^2=0$, ensuring that $\Pi^{\mathrm{ren}}(0)=0$.

\begin{equation}
    \Pi^{ren}(q^2)\equiv \Pi(q^2)-\Pi(0)
\end{equation}
The finite renormalized result simplifies to:
\begin{equation}
    \Pi^{ren}(q^2)=-\frac{g_{\chi}^2}{2\pi^2}\int^1_0 dx \text{ }x(1-x)\ln \frac{m_{\chi}^2}{m_{\chi}^2-x(1-x)q^2}
    \label{eq:eq19}
\end{equation}

For low momentum transfer $(q^2\ll 4m_{\chi}^2)$, the DM cannot be excited, so the correction is almost constant. In this limit, $\Pi^{ren}(q^2) \approx \frac{g_{\chi}^2}{60 \pi^2}\frac{q^2}{m_{\chi}^2}$, producing only a gentle shift in the propagator denominator. However, as $q^2$ approaches the threshold $(q^2 \approx 4m_{\chi}^2)$, the logarithmic term in $\Pi^{ren}(q^2)$ in eq. \ref{eq:eq19} becomes large, marking the point where virtual particles can nearly go on shell. This causes a rapid variation in the real part and the onset of an imaginary component, signaling real pair production effects. At high energies, the loop enters the pair production regime where virtual dark fermions can be created. The correction no longer grows rapidly but instead becomes logarithmic, $\Pi^{ren}(q^2) \sim \frac{g_{\chi}^2}{12\pi^2}\ln(\frac{q^2}{m_{\chi}^2})$. Physically, the vacuum starts to screen the mediator’s charge, making the effective interaction weaker and slowly varying with momentum.

Understanding the kinematic range of \(B\) meson decays in proximity to the pair production threshold $q^2 \sim 4m_\chi^2$ is of particular interest. This allows the vacuum polarization to exhibit its maximal structural variation, including mild curvature and the onset of non analytic behavior, while remaining within a perturbative and experimentally relevant regime. Consequently, all detailed fits to observables such as $P_5'$ are performed using this benchmark value, as it represents the most sensitive scenario for probing momentum dependent effects induced by the dark sector.

After incorporating the loop corrected propagator, the dark sector’s effect can be directly expressed through a modified Wilson coefficient. The relevant term, $C_9$, controls the vector interaction in the $b \to s \ell^+ \ell^-$ transition, and its shift from the SM value can encapsulate NP effects.
When a new heavy vector boson $Z'$ was introduced, the matching procedure outlined by Wolfgang Altmannshofer and David M. Straub \cite{Altmannshofer:2014rta} was followed. Since the mediator’s propagator was modified by a dark fermion loop, a vacuum polarization term was introduced. Incorporating this correction changed the momentum dependence of the interaction, meaning that the coefficient $C_9$ itself became a function of $q^2$. Using the corrected propagator from equation \ref{effprop},

\begin{align}
\Delta C_9^{\text{NP}}(q^2) &=
- \frac{\pi}{\alpha_{\text{em}} \sqrt{2} G_F V_{tb} V_{ts}^*}
\frac{g_{bs}^L \, g_\mu^V}{q^2 - M_{Z'}^2 + \Pi(q^2)}, \\[6pt]
\Delta C_{10}^{\text{NP}}(q^2) &=
- \frac{\pi}{\alpha_{\text{em}} \sqrt{2} G_F V_{tb} V_{ts}^*}
\frac{g_{bs}^L \, g_\mu^A}{q^2 - M_{Z'}^2 + \Pi(q^2)}.
\end{align}

With vector and axial couplings defined as:
\begin{equation}
g_\mu^V = g_\mu^L + g_\mu^R,
\qquad
g_\mu^A = g_\mu^L - g_\mu^R .
\end{equation}

Similarly, right handed quark currents generate contributions to the primed coefficients:
\begin{align}
\Delta C_9^{\prime\,\text{NP}}(q^2) &=
- \frac{\pi}{\alpha_{\text{em}} \sqrt{2} G_F V_{tb} V_{ts}^*}
\frac{g_{bs}^R \, g_\mu^V}{q^2 - M_{Z'}^2 + \Pi(q^2)}, \\[6pt]
\Delta C_{10}^{\prime\,\text{NP}}(q^2) &=
- \frac{\pi}{\alpha_{\text{em}} \sqrt{2} G_F V_{tb} V_{ts}^*}
\frac{g_{bs}^R \, g_\mu^A}{q^2 - M_{Z'}^2 + \Pi(q^2)}.
\end{align}

The modified coefficient was then compared with the SM baseline $C_9^{\mathrm{SM}}$ to compute the deviation $\Delta C_9$ to determine the extent to which the dark sector could account for the observed anomalies in the LHCb Collaboration angular distributions.

\section{Parameter Values} 
\label{paramValues}
A detailed justification for the benchmark parameters chosen in this study has been provided in order to ensure consistency with current experimental constraints while maintaining sensitivity to possible NP effects in the flavor sector.

The experimental measurements used in this work were obtained from the \texttt{flavio} framework, specifically from the \texttt{measurements.yml} database in which all original experimental sources were cited. Since the theoretical predictions were directly compared with the available LHCb measurements ~\cite{flavioMeasurements}, a corresponding set of dilepton invariant mass bins was defined. The analysis was performed using the following ten $q^2$ intervals:
[(0.1, 0.98),(1.1,2.5), (1.1,6.0),(2.5, 4.0), (4.0, 6.0), (6.0, 8.0), (11.0, 12.5), (15.0, 17.0), (15.0, 19.0), (17.0, 19.0)].

 Emphasis was placed on the central interval $[1.1,6.0]$ $~\mathrm{GeV}^2$, corresponding to the region in which the most statistically significant deviations from SM expectations had been reported in the June 13, 2025 analysis of the $P_5'$ angular observable. Furthermore, the loop induced vacuum polarization correction was evaluated within this range in order to ensure that the condition $q^2 < 4m_\chi^2$ remained satisfied, thereby keeping the polarization function real and physically meaningful.

The mediator mass $m_{Z'}$ was chosen to remain compatible with present collider limits while still allowing measurable effects in low energy flavor observables. Analyses excluded sequential SM like $Z'$ bosons with masses below approximately $5$--$5.5~\mathrm{TeV}$ under the assumption of dominant decays into charged leptons \cite{ATLAS:2019dilepton,CMS:2021dilepton}. In the present framework, however, invisible decay channels of the form $Z' \to \chi\bar{\chi}$ were permitted, significantly suppressing the visible dilepton branching fraction and consequently relaxing the collider constraints \cite{Arcadi:2018DMspectrum}. A benchmark value of $m_{Z'} = 3~\mathrm{TeV}$ was therefore adopted, while an extended scan over the range $1~\mathrm{TeV} \leq m_{Z'} \leq 3~\mathrm{TeV} $ was also performed. Within this interval, the dimuon production cross section remained below existing experimental limits for moderate values of the leptonic coupling $ g_{\mu} \lesssim 0.3 $, while the mediator remained sufficiently light for loop level contributions capable of generating $\Delta C_9 \sim$ -1 to be produced, as preferred by global fits to the $B \to K^*\mu^+\mu^-$ data \cite{Altmannshofer:2021C9fit,Capdevila:2017bsm}.

The dark Dirac fermion mass $m_\chi$ was constrained by the requirement that the vacuum polarization correction remain real. In order to avoid imaginary contributions arising from on shell dark fermion pair production, the condition $q_{\mathrm{max}}^2 < 4m_\chi^2 $ was imposed. Since the central $q^2$ region extended up to approximately $q^2 \simeq 6~\mathrm{GeV}^2$, a lower bound of $m_\chi \gtrsim 1.22~\mathrm{GeV}$ was obtained.

To systematically explore the impact of dark sector vacuum polarization, the loop correction was evaluated across three mass regimes: \(m_\chi \ll 1\,\mathrm{GeV}\) (light), \(m_\chi \sim 1\,\mathrm{GeV}\) (threshold), and \(m_\chi \gg 1\,\mathrm{GeV}\) (heavy).

\begin{table*}[t]
 \caption{NP parameters of the LFU-V $Z'$ model, the scan ranges explored in this work, and the corresponding phenomenological considerations.}
\centering
\begin{tabular}{l c c c l}
\hline
\textbf{Parameter} & \textbf{Range} & \textbf{Units} & \textbf{Varied in} & \textbf{Parameter Name} \\
\hline

$m_{Z'}$ 
& $[1,3]$ 
& TeV 
& All models 
& Mass of the $Z'$ propagator \\

$g^{L}_{bs}$ 
& $[-0.075,-0.005]$ 
& -- 
& All models 
& Flavour violating quark coupling (Left handed current) \\

$g^{R}_{bs}$ 
& $[-0.075,-0.005]$ 
& -- 
& LRA, LRS 
& Flavour violating quark coupling (Right handed current) \\

$g^{L}_{\mu}$ 
& $[0.025,0.250]$ 
& -- 
& All models 
& Muonic coupling (Left handed current) \\

$g^{R}_{\mu}$ 
& $[0.025,0.250]$ 
& -- 
& LRA, LRS 
& Muonic coupling (Right handed current) \\

$g_{\chi}$ 
& $[0.2,1.2]$ 
& -- 
& All models 
& Dark sector coupling \\

$m_{\chi}$ 
& $2$ 
& GeV 
& All models 
& Mass of the dark matter fermion \\
\hline
\end{tabular}
\label{tab:model_parameters}
\end{table*}

The full set of NP parameters, the ranges scanned in each chiral realization, and the associated phenomenological constraints are summarized in Table~\ref{tab:model_parameters}. The coupling intervals used in the grid are listed for convenience in Table~\ref{tab:couplings}.

\begin{table}[ht]
\caption{Coupling scan ranges used in the parameter space exploration.}

\centering
\begin{tabular}{c c}
\hline
Coupling & Range \\
\hline
$g^{L,R}_{bs}$ & $[-0.075,\,-0.005]$ \\
$g^{L,R}_\mu$  & $[0.025,\,0.250]$ \\
$g_\chi$ & $[0.20,\,1.20]$ \\
\hline
\end{tabular}
\label{tab:couplings}
\end{table}

The flavor changing coupling $g_{bs}$ was taken to govern the strength of the $b \to s$ transition. Since this coupling was tightly constrained by measurements of $B_s-\bar{B}_s$ mixing \cite{DiLuzio:2019Bsmixing}, only a restricted parameter interval was considered. The negative sign of the coupling was chosen in order to ensure destructive interference with the SM amplitude, as preferred by global fits to the $P_5'$ anomaly. The leptonic coupling $g_{\mu}$ was varied within a range consistent with constraints derived from the anomalous magnetic moment of the muon \cite{Aoyama:2020gminus2} and from LHC dilepton searches \cite{CMS:2021dilepton}. At the same time, the selected interval was required to remain sufficiently large to generate an observable modification in the transition amplitude. The dark sector coupling $g_{\chi}$ was chosen such that larger values enhanced the invisible decay channel $Z' \to \chi \bar{\chi}$,
through which visible dilepton signatures were suppressed and collider bounds on $m_{Z'}$ were effectively relaxed \cite{Arcadi:2018DMspectrum}. The selected range additionally satisfied the perturbativity condition as:
\[\frac{g_{\chi}^2}{4\pi} < 1,\] 
while allowing the dark fermion loop to produce a significant modification of the propagator structure.

\section{Computational Framework}
\label{compFW}

The computational analysis was carried out using \texttt{flavio}~\cite{straub2018flavio}, through which the theoretical predictions within the present framework were combined with experimental measurements to quantify the level of agreement between theory and data. The experimental measurements of the $P_5'$ angular observable, together with their associated uncertainties, were extracted directly from the internal database implemented in the \texttt{flavio} framework~\cite{flavioMeasurements}, which contains the measurements from the LHCb 2020 analysis of the $P$-wave angular observables. The corresponding theoretical predictions for $P_5'$ were calculated in the same $q^2$ intervals, allowing a direct comparison between the theoretical predictions and the experimental measurements on a bin by bin basis. The analysis was performed through a systematic study of several chiral benchmark realizations of the $Z'$ mediator, allowing different Lorentz structures of the new interaction to be tested against the experimental data. Rather than restricting the analysis to a single effective field theory configuration, three complementary benchmark scenarios were considered according to the chirality of the $Z'$ couplings.

The analysis begins with the left handed scenario (LHS), which closely matches the SM chiral structure. As a benchmark, only the left handed quark and lepton couplings are retained, while all right handed couplings are set to zero.
\[
\mathcal L^{int}_{LHS} \supset g_{bs}^L \bar s\gamma^\mu P_LbZ'_\mu + g_\mu^L \bar \mu\gamma^\mu P_L\mu Z'_\mu.
\]
Within this framework, the right handed couplings satisfied
\[
g_{bs}^R = g_\mu^R = 0.
\]
This realization mirrored the chiral structure of the SM weak interaction and therefore served as a minimal benchmark for NP effects. The dominant loop induced contribution was generated in the Wilson coefficient $C_9$, while contributions to $C_{10}$ remained correlated through
\[
\Delta C_9 = - \Delta C_{10} \qquad \Delta C_9' = \Delta C_{10}' = 0
\]

The corresponding vector and axial couplings were obtained as
\[
g_{bs,\mu}^{V}=g_{bs,\mu}^{A}=g_{bs,\mu}^{L}.
\]
A four dimensional grid scan consisting of 6561 parameter combinations was performed,
\[
(m_{Z'},m_\chi,g_{bs}^{L},g_{\mu}^{L},g_{\chi})=(9,1,9,9,9),
\]
in order to determine the best fit parameter region.

The introduction of a right handed coupling in the $Z'$ is motivated by analyses which report that the measured value of the observable $BR(B_s \to \mu^+ \mu^-)$ closely matches the SM prediction (further elaborated in Section \ref{sec:bsmumuanalysis}).

In the left right asymmetric scenario (LRA), the left- and right-handed couplings were treated as independent free parameters,
\[
\mathcal L^{int}_{LRA} \supset \bar s\gamma^\mu(g_{bs}^LP_L+g_{bs}^RP_R)bZ'_\mu +
\bar \mu \gamma^\mu(g_\mu^LP_L + g_\mu^R P_R)\mu Z'_\mu.
\]
This realization represented the most general framework considered in the present study and allowed simultaneous vector and axial vector contributions to be generated. Consequently, correlated shifts in both $C_9$ and $C_{10}$ were produced. The effective vector and axial couplings were expressed as
\[
g_{bs,\mu}^{V}=g_{bs,\mu}^{L}+g_{bs,\mu}^{R},
\]
\[
g_{bs,\mu}^{A}=g_{bs,\mu}^{L}-g_{bs,\mu}^{R}.
\]
A six dimensional grid scan containing 21875 parameter combinations was evaluated,
\[
(m_{Z'},m_\chi,g_{bs}^{L},g_{bs}^{R},g_{\mu}^{L},g_{\mu}^{R},g_{\chi})
=(7,1,5,5,5,5,5),
\]
thereby allowing the full Lorentz structure of the interaction to be explored.

In the left right symmetric Scenario (LRS), equal left- and right-handed couplings were imposed,
\[
\mathcal L^{int}_{LRS} \supset g_{bs}\bar s \gamma^\mu b Z'_\mu
+ g_\mu \bar \mu \gamma^\mu Z'_\mu.
\]
Within this configuration,
\[
g_{bs}^{L}=g_{bs}^{R},
\qquad
g_{\mu}^{L}=g_{\mu}^{R},
\]
yielding a purely vector like interaction structure. As a consequence, the axial contribution vanished, $g_{bs,\mu}^{A}=0$, while the dominant NP effect appeared in the vector Wilson coefficient,
\[
\Delta C_9 \neq 0,
\qquad
\Delta C_{10}=0.
\]
The effective vector coupling was obtained as
\[
g_{bs,\mu}^{V}=g_{bs,\mu}^{L}+g_{bs,\mu}^{R}.
\]
A four dimensional scan consisting of 6561 parameter combinations was carried out,
\[
(m_{Z'},m_\chi,g_{bs},g_{\mu},g_{\chi})=(9,1,9,9,9),
\]
in order to determine whether a purely vector interaction could reproduce the observed angular anomalies.

Although the parameter scan was optimized primarily using the $P_5'$ observable, the resulting best fit regions were subsequently evaluated against additional observables in the $B \to K^* \mu^+ \mu^-$ and $B_s \to \mu^+ \mu^-$ decay channels in order to perform consistency checks. Through this procedure, the preferred parameter configurations were required to remain compatible with the broader set of flavor physics constraints.

\begin{figure*}[thbp]
\centering
    \includegraphics[width=0.7\linewidth]{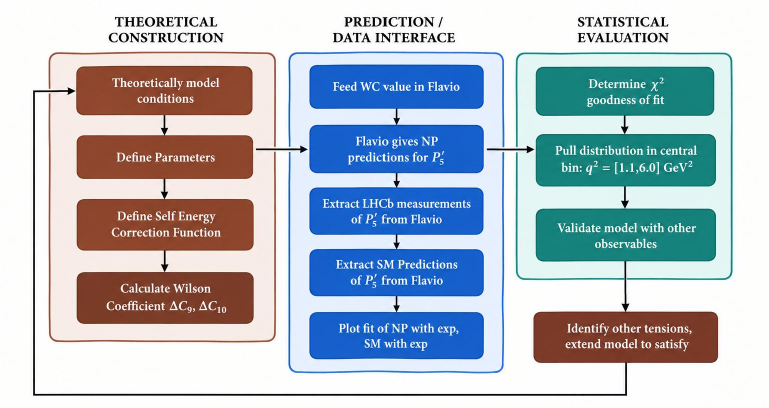}
\caption{Schematic of the analysis workflow showing theoretical construction of the model and Wilson coefficients on the left, interfacing with \texttt{flavio} for $P_5'$ predictions plus data in the centre, followed by statistical evaluation with iterative model refinement on the right.}
\label{fig:workflow}
\end{figure*}

The analysis chain, from the construction of $\Delta C_9(q^2)$ through the \texttt{flavio} interface to the $\chi^2$ evaluation, is summarized in Fig.~\ref{fig:workflow}.

The agreement between theoretical predictions and experimental measurements was quantified using the $\chi^2$ statistic, which served as the primary goodness of fit estimator,
\begin{equation}
    \chi^2 = \sum_i
    \left(
    \frac{
    P_{5,\mathrm{theory}}'^{(i)}
    -
    P_{5,\mathrm{exp}}'^{(i)}
    }{
    \sigma_{\mathrm{exp}}^{(i)}
    }
    \right)^2.
\end{equation}
Here,
$P_{5,\mathrm{theory}}'^{(i)}$
denoted the theoretical prediction in the $i^{\mathrm{th}}$ $q^2$ bin,
$P_{5,\mathrm{exp}}'^{(i)}$
represented the corresponding experimental central value, and
$\sigma_{\mathrm{exp}}^{(i)}$
corresponded to the experimental uncertainty. Smaller values of $\chi^2$ indicated improved agreement between theory and experiment.

In order to obtain a scale independent quantity, the total $\chi^2$ value was normalized by the number of degrees of freedom,
\begin{equation}
    \frac{\chi^2}{\mathrm{dof}}
    =
    \frac{1}{N_{\mathrm{bins}}-N_{\mathrm{params}}}\chi^2.
\end{equation}

The level of agreement was additionally characterized through the pull variable,
\begin{equation}
    \mathrm{Pull}^{(i)}
    =
    \frac{
    P_{5,\mathrm{theory}}'^{(i)}
    -
    P_{5,\mathrm{exp}}'^{(i)}
    }{
    \sigma_{\mathrm{exp}}^{(i)}
    }.
\end{equation}
The pull distribution provided an intuitive measure of the deviation between theoretical predictions and experimental data in units of the experimental uncertainty. Pull values centered near zero with magnitudes satisfying
$|\sigma| \lesssim 1$
were interpreted as indicating consistency with the measured data within uncertainties.

The vacuum polarization contribution generated by the dark fermion loop entered as a subleading correction to the $Z'$ propagator and therefore remained parametrically smaller than the tree level exchange contribution. Consequently, its primary impact was manifested not through a large overall shift in the Wilson coefficients, but rather through the introduction of a nontrivial momentum dependence.

In order to isolate this effect, the loop induced contribution was analyzed relative to the corresponding tree level prediction. The total correction to the Wilson coefficient was decomposed as:
$ \Delta C_9(q^2) = \Delta C_9^{\mathrm{tree}}+ \delta C_9^{\mathrm{loop}}(q^2)$,
where
$\Delta C_9^{\mathrm{tree}}$
denoted the momentum independent tree level contribution and
$\Delta C_9^{\mathrm{loop}}(q^2)$
represented the vacuum polarization correction generated by the dark sector loop.

After subtraction of the constant tree level component, the residual quantity
\begin{equation}
    \delta_{\mathrm{loop}}(q^2)
    =
    \Delta C_9(q^2)
    -
    \Delta C_9^{\mathrm{tree}}
\end{equation}
was obtained, directly probing the momentum dependence induced by the dark fermion loop.

\section{Results }
\label{results}
Extensive grid scans were performed across the three chiral realizations of the model by testing thousands of parameter combinations in order to identify regions of parameter space yielding optimal agreement with experimental data from the  measurements of $B \to K^* \mu^+ \mu^-$. 
In total, 34997 benchmark points were evaluated: 6561 in the LHS, 21875 in the LRA, and 6561 in the LRS.

\subsection{Fits to $P_5'$}
\begin{figure}[htbp]
    \centering
    \includegraphics[width=0.7\linewidth]{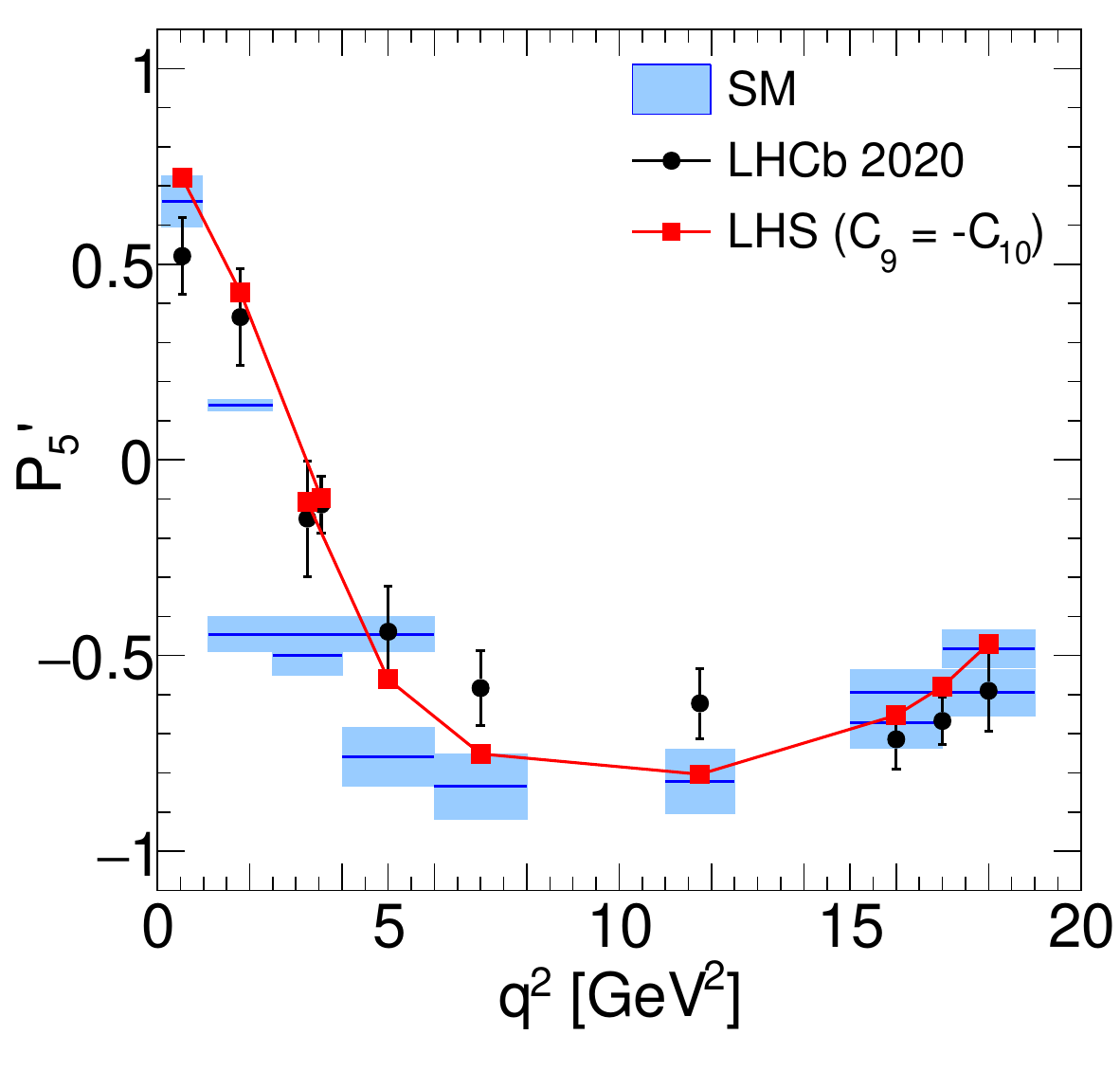}
    \caption{$P_5'$ fits for the left handed scenario (LHS) ($C_9 = -C_{10}$) and SM compared against LHCb 2020 data. The LHS fit provides a substantial improvement over the SM baseline}
    \label{fig:lhs}
\end{figure}

\begin{figure}[ht]
    \centering
    \includegraphics[width=0.7\linewidth]{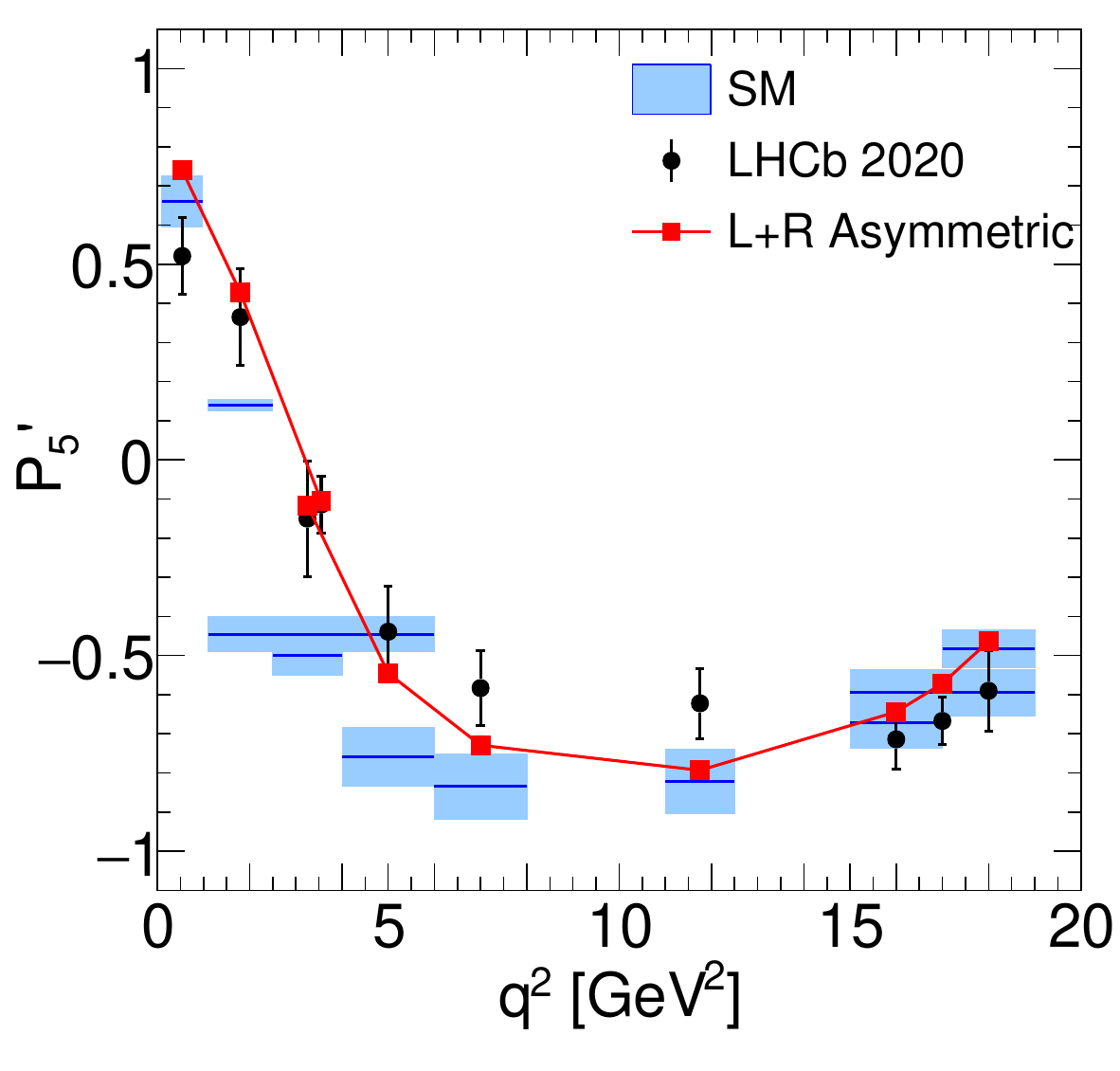}
    \caption{$P_5'$ fits for the left right asymmetric (LRA) realization alongside SM predictions and LHCb 2020 measurements. By allowing chirally asymmetric vector dominant interactions, the LRA fit better tracks the experimental data points across low  and mid $q^2$ bins compared to the SM.}
    \label{fig:lra}
\end{figure}

\begin{figure}[ht]
    \centering
    \includegraphics[width=0.7\linewidth]{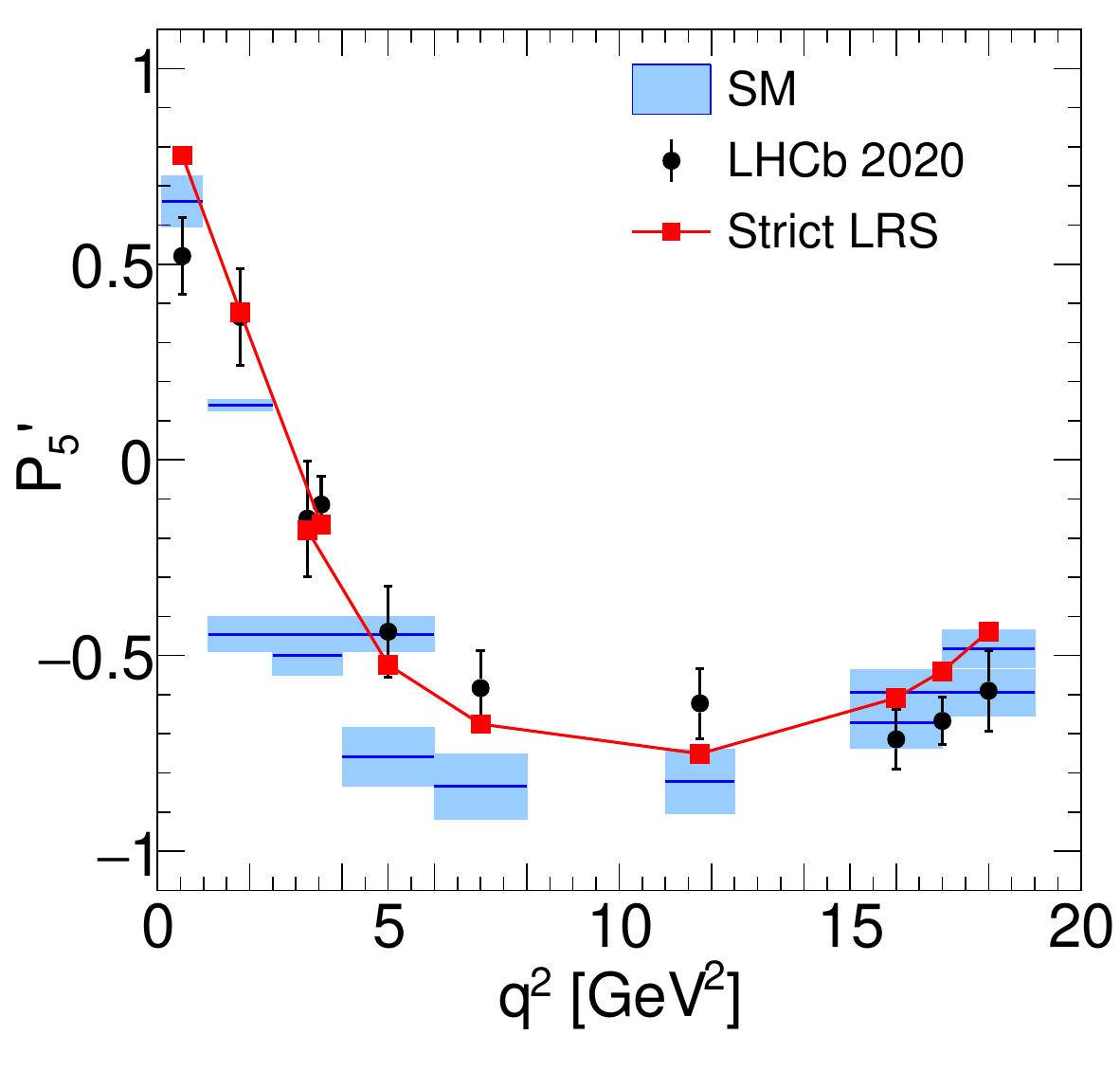}
    \caption{$P_5'$ fits for the Strict left right Symmetric (LRS) scenario compared to the SM and LHCb data. Due to the rigid symmetry constraint imposing equal chiral contributions, the LRS realization offers a noticeably poorer fit to the experimental $P_5'$ data compared to its asymmetric counterparts.}
    \label{fig:lrs}
\end{figure}

The resulting fits to the $P'_5$ observable for the three benchmark realizations are presented in Figs.~\ref{fig:lhs}--\ref{fig:lrs}. Significant improvement relative to the SM prediction was obtained in all three scenarios, particularly in the central $q^2$ region where the experimental anomaly is most pronounced.

The purely left handed interaction structure considered in the LHS realization was found to provide substantial improvement over the SM expectation, which exhibits an approximately $3\sigma$ deviation in the central $q^2$ bin. The tension was reduced by nearly $92\%$, yielding a pull value of
$0.24\sigma$.

In the LRA realization, the interaction structure was found to be vector dominant but chirally asymmetric. The quark sector remained predominantly left handed,
\[
g_{bs}^R/g_{bs}^L \approx 0.087,
\]
while the lepton sector exhibited a strong right handed preference,
\[
g_\mu^R/g_\mu^L = 10.
\]
This configuration yielded the strongest agreement with experimental data, corresponding to a pull value of
$0.12\sigma$.
\\

In the LRS realization, equal left- and right-handed couplings were imposed. Although the fit remained phenomenologically viable, the resulting agreement was slightly weaker than in the LHS and LRA configurations, yielding a pull value of $-0.71\sigma$ with $\chi^2 = 19.46$.

The corresponding best fit parameter configurations obtained in each benchmark realization are summarized in Table~\ref{tab:bestfit_summary}.

\begin{table}[ht]
\centering
\setlength{\tabcolsep}{8pt}
\renewcommand{\arraystretch}{1.2}

\begin{tabular}{c c c c}
\hline
Parameter & LHS & LRA & LRS \\
\hline

$m_{Z'}$ (GeV)
& 1250
& 3000
& 3000 \\

$m_\chi$ (GeV)
& 2
& 2
& 2 \\

$g_{bs}^{L}$
& $-0.0575$
& $-0.0575$
& -- \\

$g_{bs}^{R}$
& --
& $-0.0050$
& -- \\

$g_{bs}^{L,R}$
& --
& --
& $-0.0663$ \\

$g_{\mu}^{L}$
& $+0.0531$
& $+0.0250$
& -- \\

$g_{\mu}^{R}$
& --
& $+0.2500$
& -- \\

$g_{\mu}^{L,R}$
& --
& --
& $+0.0531$ \\

$g_\chi$
& $0.2000$
& $0.2000$
& $0.2000$ \\

Pull $[1.1,6.0]$
& $0.24\sigma$
& $0.12\sigma$
& $-0.71\sigma$ \\

$\chi^2$
& 16.99
& 17.01
& 19.46 \\

\hline
\end{tabular}

\caption{Best fit parameter configurations and goodness of fit values obtained in the three benchmark realizations.}
\label{tab:bestfit_summary}

\end{table}

Overall, all three benchmark realizations were found to provide viable parameter regions capable of significantly alleviating the $P'_5$ anomaly observed in the LHCb data. The LHS and LRA realizations yielded the strongest quantitative agreement with experiment, while the fully symmetric realization remained compatible with the measured observables, although slightly less optimal in terms of the overall goodness of fit.

To quantify the impact of dark sector vacuum polarization on the effective Wilson coefficient $C_9$, the loop induced correction $\Pi(q^2)$ and its propagation into $\Delta C_9(q^2)$ are analyzed. In particular, it is investigated whether vacuum polarization effects generated by a Dirac dark fermion $\chi$ can induce observable distortions in the angular observables of $B \to K^* \mu^+ \mu^-$. 

\begin{figure}[htbp]
\centering
    \includegraphics[width=0.45\linewidth]{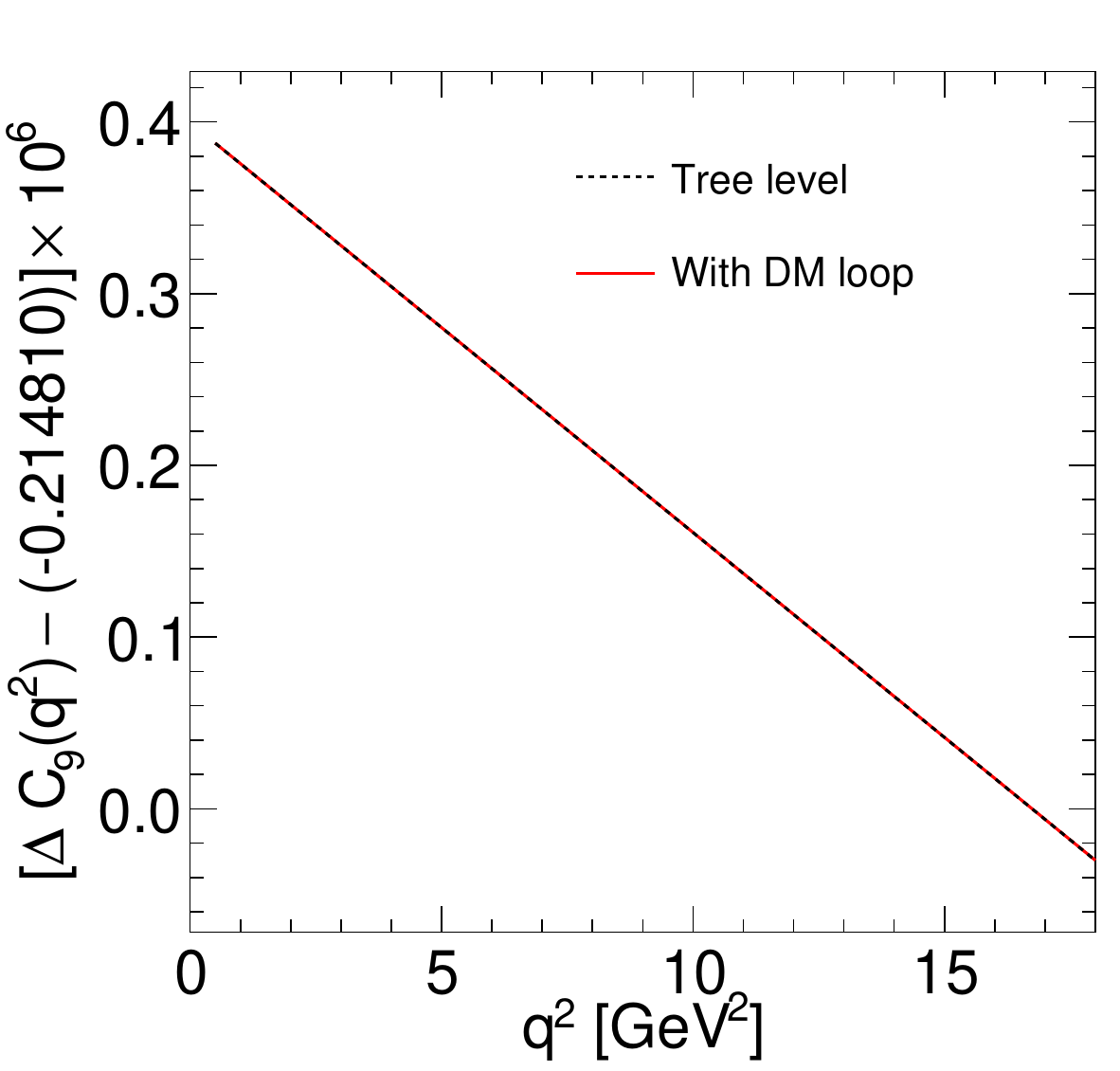}
     \includegraphics[width=0.45\linewidth]{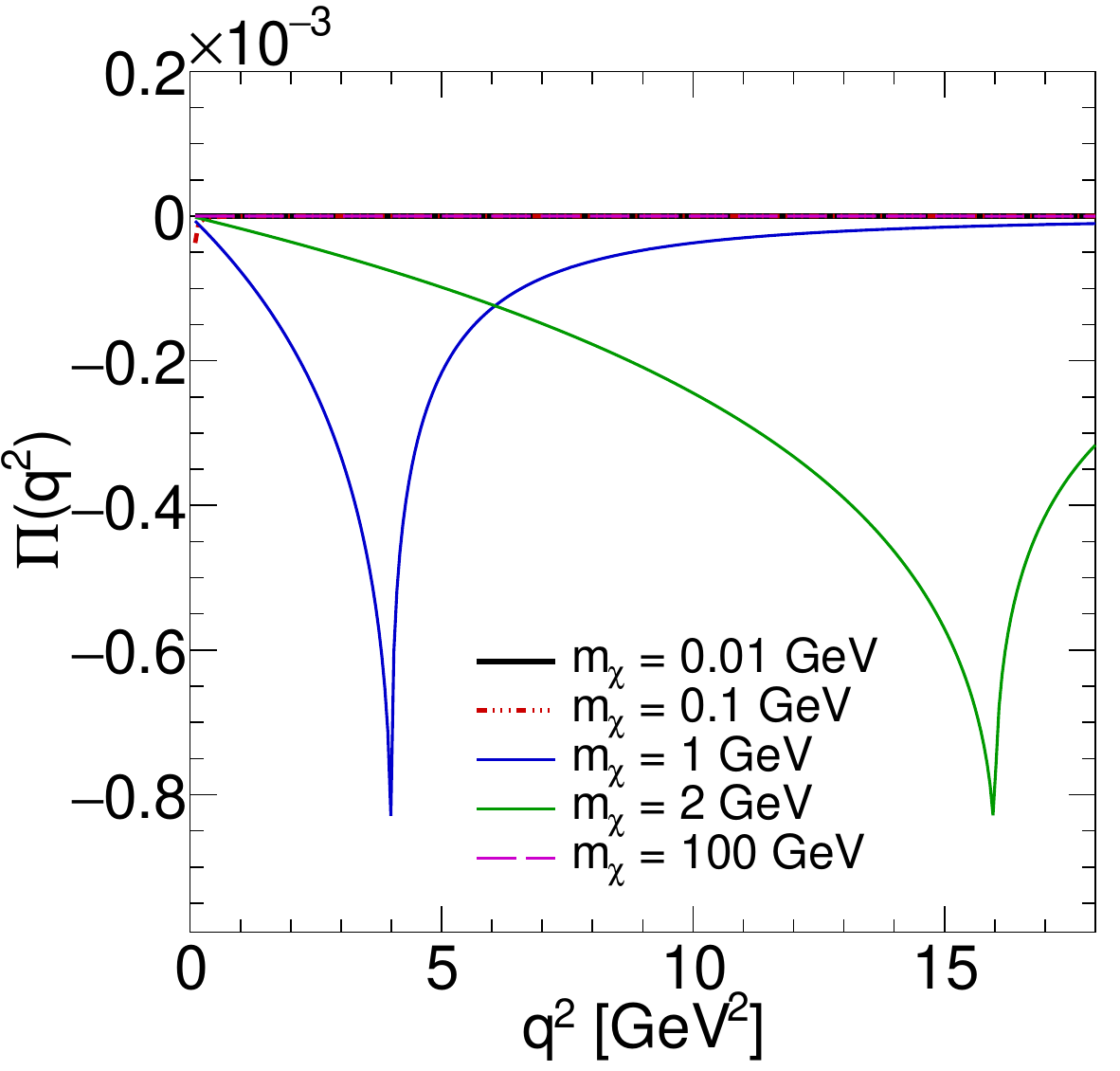}
       \caption{Comparison of tree level and loop corrected $\Delta C_9(q^2)$ contributions Dependence of the vacuum polarization function $\Pi(q^2)$ on the DM mass scale. Impact of dark sector vacuum polarization on the effective Wilson coefficient.}
\label{fig:vp_effects}
\end{figure}

To assess the phenomenological significance of these corrections, the tree level and loop corrected contributions to $\Delta C_9(q^2)$ are compared, as shown in Fig.~\ref{fig:vp_effects}.

Although Fig.~\ref{fig:vp_effects} demonstrates a non trivial momentum dependence in the vacuum polarization function, the resulting loop induced modification to $\Delta C_9(q^2)$ remains extremely small. Consequently, vacuum polarization effects do not significantly alter the overall normalization of $C_9$, but instead generate only a smooth and subleading deformation.

\begin{figure}
\centering
    \includegraphics[width=\linewidth]{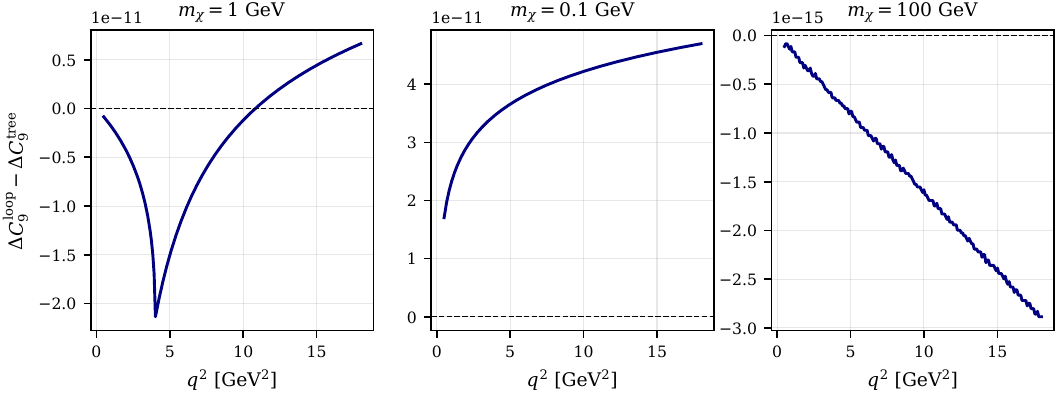}
    \caption{Loop induced deviation in the scenario of $m_\chi = 1$ GeV, $m_\chi = 100$ MeV, $m_\chi = 100$ GeV respectively. A sharp resonance like dip near $q^2 \approx 4\text{ GeV}^2$ in the first panel is observed. Middle panel shows a smooth, monotonic rise across the full $q^2$ spectrum. In the right panel the heavy mass scale suppresses the loop integral and causes a continuous downward trend with $q^2$ (the fine stair step jaggedness is a non physical numerical discretization artifact.)}
\label{fig:loop_grid_mass}
\end{figure}

\begin{figure}
\centering
    \includegraphics[width=\linewidth]{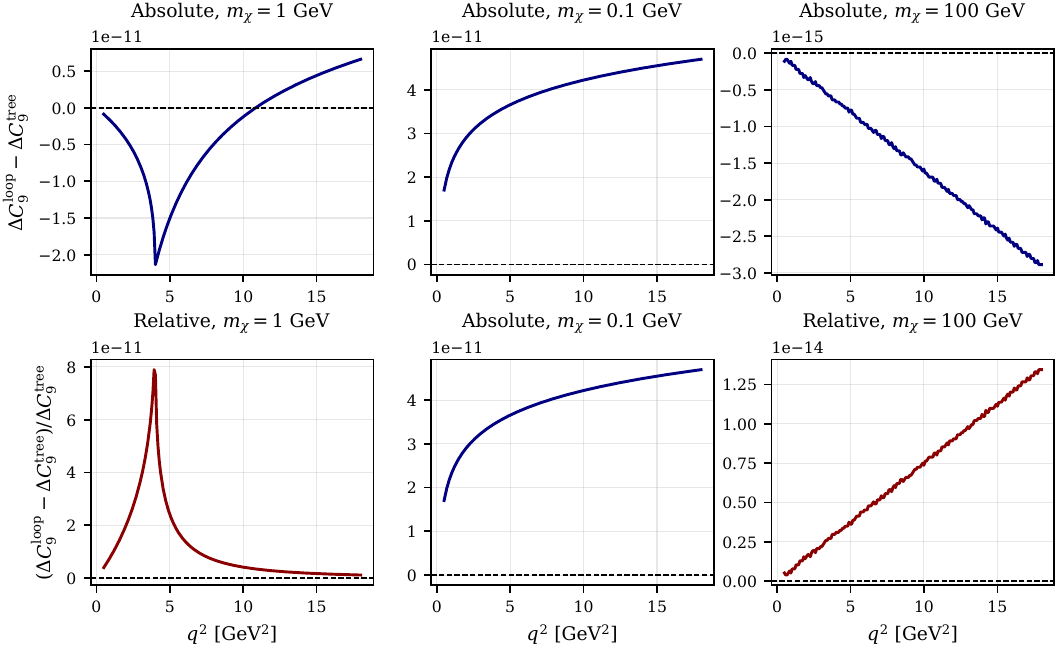}
    \caption{Relative deviation ($m_\chi = 1$ GeV, $m_\chi = 100$ MeV and $m_\chi = 100$ GeV) in $\Delta C_9(q^2)$ for representative DM masses. These panels illustrate the corresponding relative momentum dependence.}
\label{fig:loop_grid1}
\end{figure}

To isolate the contribution arising purely from the dark sector loop, the loop induced shift is defined as
\begin{equation}
\delta_{\text{loop}}(q^2) = \Delta C_9^{\text{loop}}(q^2) - \Delta C_9^{\text{tree}}.
\end{equation}

To quantify the relative size of the correction, the normalized deviation  is further defined as 
\begin{equation}
\delta_{\text{rel}}(q^2) =
\frac{
\Delta C_9^{\text{loop}}(q^2) - \Delta C_9^{\text{tree}}
}{
\Delta C_9
}.
\end{equation}
The absolute loop induced shift $\delta_{\mathrm{loop}}(q^2)$ for three representative dark fermion masses is shown in Fig.~\ref{fig:loop_grid_mass}. The corresponding relative correction $\delta_{\mathrm{rel}}(q^2)$ is shown in Fig.~\ref{fig:loop_grid1}. After subtracting the overall constant tree level contribution, a residual non trivial $q^2$ dependence becomes visible.

For light DM masses ($m_\chi \sim 1$ GeV), the correction exhibits a pronounced variation near the kinematic threshold
\begin{equation}
q^2 \simeq 4m_\chi^2,
\end{equation}
which manifests as a characteristic curvature in $\delta_{\text{loop}}(q^2)$. This threshold behavior originates from the opening of the virtual dark fermion pair contribution inside the vacuum polarization loop.

For intermediate masses, the deviation remains smooth but increasingly suppressed. In the heavy mass regime ($m_{\chi} \gtrsim 100 ~\mathrm{GeV}$), the correction becomes nearly linear and featureless, reflecting the decoupling behavior of the heavy dark sector. 

Although the absolute magnitude of the loop correction remained phenomenologically subleading over most of the parameter space, its momentum dependence became enhanced near kinematic thresholds and persisted even in regions where the overall shift in $C_9$ remained approximately constant. Consequently, the energy dependence of the correction, rather than its absolute normalization, was identified as the primary observable signature of the vacuum polarization effect. Variations in the curvature and shape of
$\delta_{\mathrm{loop}}(q^2)$ across the accessible kinematic region therefore provided a physically meaningful discriminator between purely tree level explanations and loop induced dark sector dynamics.

\subsection{Consistency check with other observables}
 While the parameter scans were optimized using $P_5'$, the best fit regions were subsequently validated against complementary observables in $B \to K^* \mu^+ \mu^-$, $B \to  \phi \mu^+ \mu^-$, and $B_s \to \mu^+\mu^-$ decays. This ensures the preferred parameter configurations do not improve $P_5'$ at the expense of other measured quantities.

\subsubsection{$B_s \to \mu^+ \mu^-$}
\label{sec:bsmumuanalysis}

\begin{figure}[htbp]
    \centering
    \includegraphics[width=0.6\linewidth]{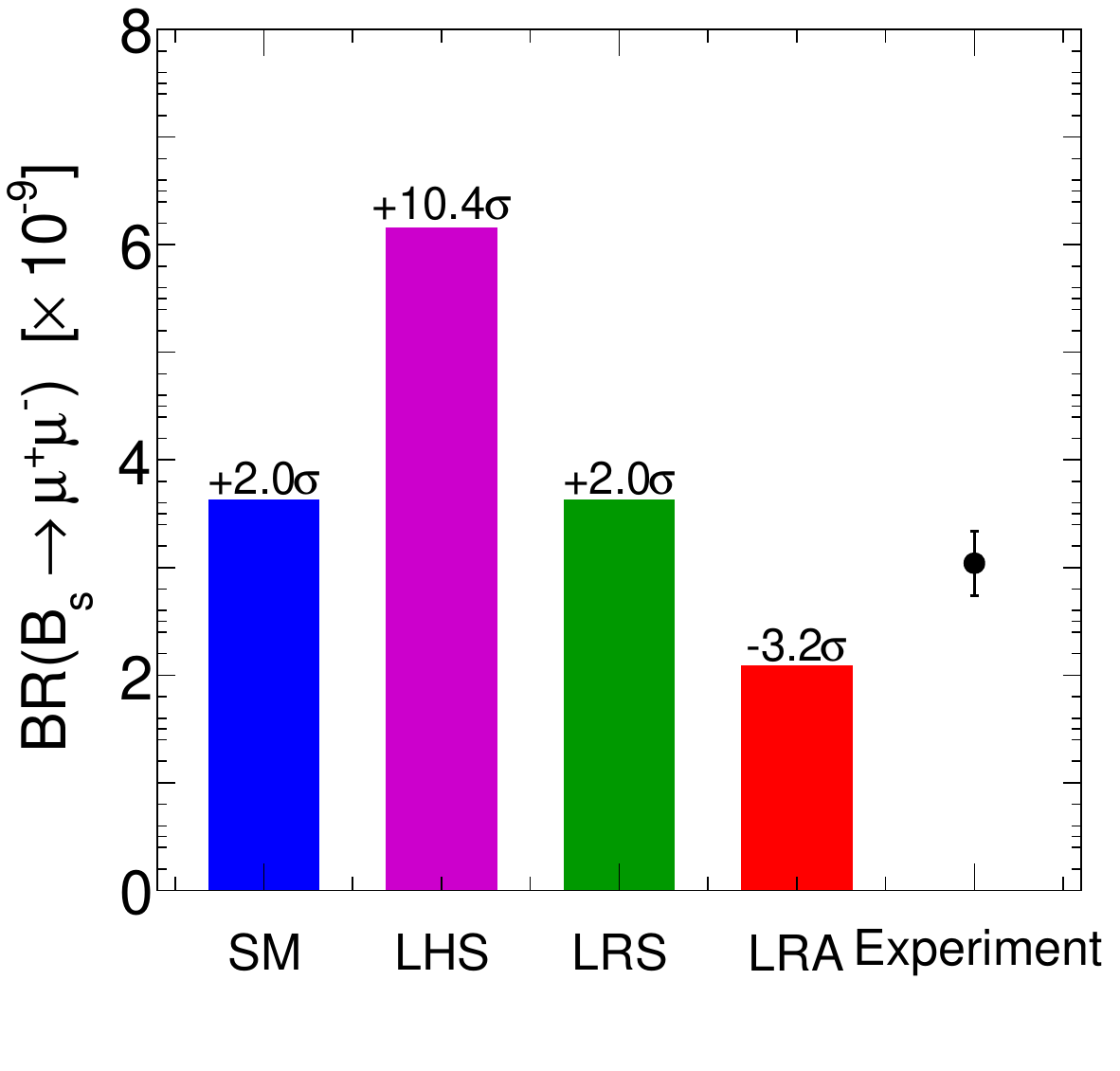}
    \caption{Consistency of SM, LHS, LRS, and LRA predictions with experimental measurements of the $B_s \to \mu^+\mu^-$ branching ratio. While the SM and LRS models remain consistent with experimental bounds, the LHS model exhibits severe tension due to overpredicting the branching ratio, and the LRA model underestimates the rate.}
    \label{fig:bsmumu}
\end{figure}
The comparison of the SM, LHS, LRA, and LRS predictions with the measured $\mathrm{BR}(B_s\to\mu^+\mu^-)$ is shown in Fig.~\ref{fig:bsmumu}.
The purely leptonic decay of a \(B\) meson with a strange quark to a muon antimuon pair is essential in determining the chirality of the Z', because its branching ratio is largely in coherence with the SM. $BR(B_s \to \mu^+\mu^-)$ provides a stringent test of the axial Wilson coefficient structure, because it is largely dominated by the operator $O_{10}$. As a result, its branching ratio is directly sensitive to the Wilson coefficient $C_{10}$. This implies that viable extensions must satisfy $\Delta C_{10} \approx 0$. This was the primary motive for exploring the LRS model, since it results in $\Delta C_{10} = 0$. 

\subsubsection{Angular Observables of $B \to K^* \mu^+ \mu^-$}
\begin{figure}
\centering
    \includegraphics[width=1.0\linewidth]{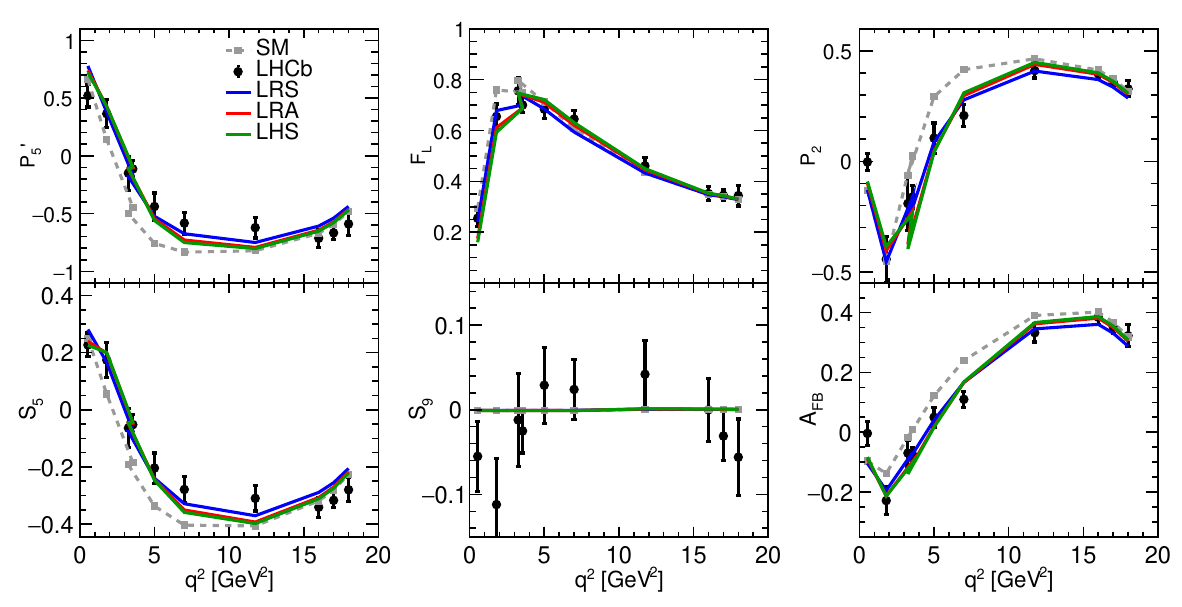}
   \caption{$P'_5$, $F_L$, $A_{FB}$, $P_2$, $S_5$, $S_9$, Comparison of angular observables across models in $B \to K^*\mu^+\mu^-$ decays.}
   \label{fig:angular-obs}
\end{figure}

The LRA, LHS and LRS best fit points were tested against the full suite of angular observables: $F_L$ (longitudinal polarization fraction, sensitive to $C_7-C_9$ interplay), $A_{FB}$ (forward backward asymmetry, probing vector axial interference), $S_5$ (unnormalized $P_5'$ analogue), $S_9$ (sensitive to right handed currents via primed operators), and $P_2$ (secondary optimized observable). The comparison is shown in Fig.~\ref{fig:angular-obs}.

Across all three scenarios, the models provide a consistently good description of the data for $F_L, A_{FB}, P_2, S_5$, with predicted central values closely tracking experimental measurements across the full $q^2$ range. In particular, the improvement in $P_5'$ relative to the SM is preserved while maintaining compatibility with other observables, indicating that the fitted Wilson coefficient structure does not introduce tensions elsewhere in the angular spectrum.

\begin{figure}
    \centering
    \includegraphics[width=\linewidth]{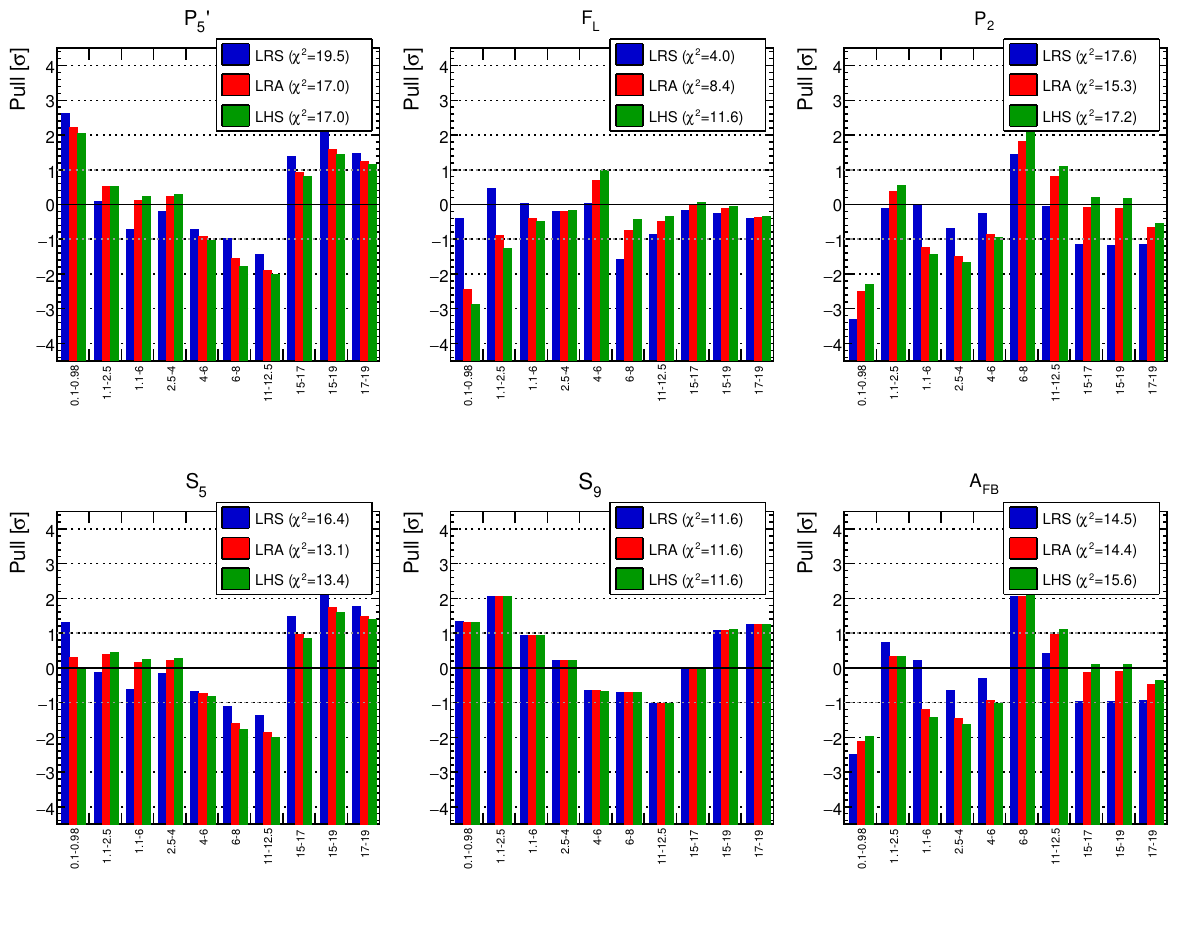}
    \caption{Pull distribution comparison across all $q^2$ bins for all models and observables.}
    \label{fig:pulldistr}
\end{figure}
\begin{figure}
    \centering
    \includegraphics[width=\linewidth]{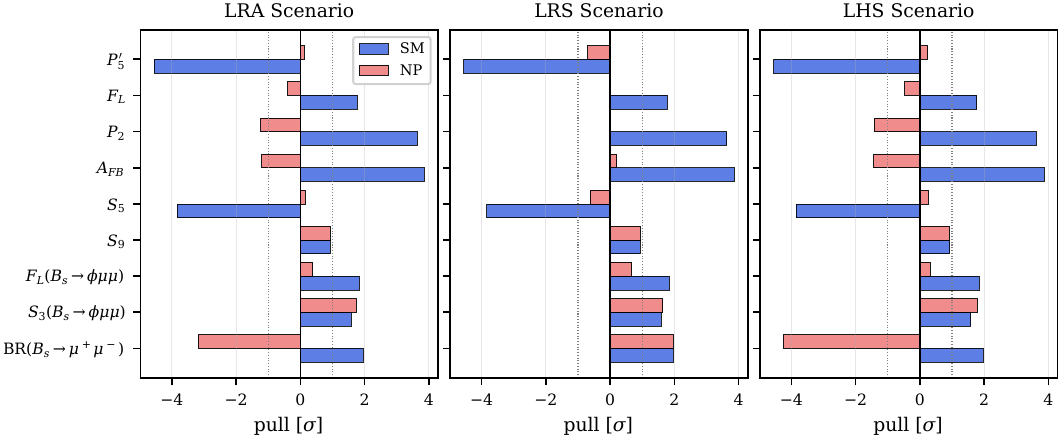}
    \caption{Pull distribution comparison in central $q^2$ bin $[1.1, 6.0] GeV^2$ for all models and observables.}
    \label{fig:pulls-central}
\end{figure}
The pull in every $q^2$ bin is shown in Fig.~\ref{fig:pulldistr}. The same comparison restricted to the central bin $[1.1,6.0]\,\mathrm{GeV}^2$ is shown in Fig.~\ref{fig:pulls-central}.

Among the three scenarios considered, the left right asymmetric (LRA) configuration provides the best fit, with $\chi^2/\text{dof} = 1.329$, followed by the left right symmetric (LRS) case with $\chi^2/\text{dof} = 1.392$. The purely LHS yields the weakest agreement, with $\chi^2/\text{dof} = 1.439$.

Although all three models achieve reasonable agreement with data, the improvement in the LRA scenario indicates that allowing independent left- and right-handed couplings introduces additional flexibility in fitting the angular distributions. In contrast, the LHS scenario, while capable of addressing specific observables such as $P'_5$, is comparatively more constrained when evaluated against the full set of measurements. 

The preference for the LRA scenario suggests that purely left handed interactions may be insufficient to simultaneously accommodate all observables, and that a mild breaking of left right symmetry can improve global consistency. We study the optimal degree of asymmetry for coupling strengths in an LRA model. 

\subsubsection{$B_s \to \phi\mu^+ \mu^-$}
As an additional consistency check, we examine angular observables in the decay $B_s \to \phi \mu^+ \mu^-$. This channel provides an independent probe of the same underlying quark level transition $b \to s \mu^+ \mu^-$, but with different hadronic form factors and sensitivities to operator combinations.

\begin{figure}
\centering
    \includegraphics[width=\linewidth]{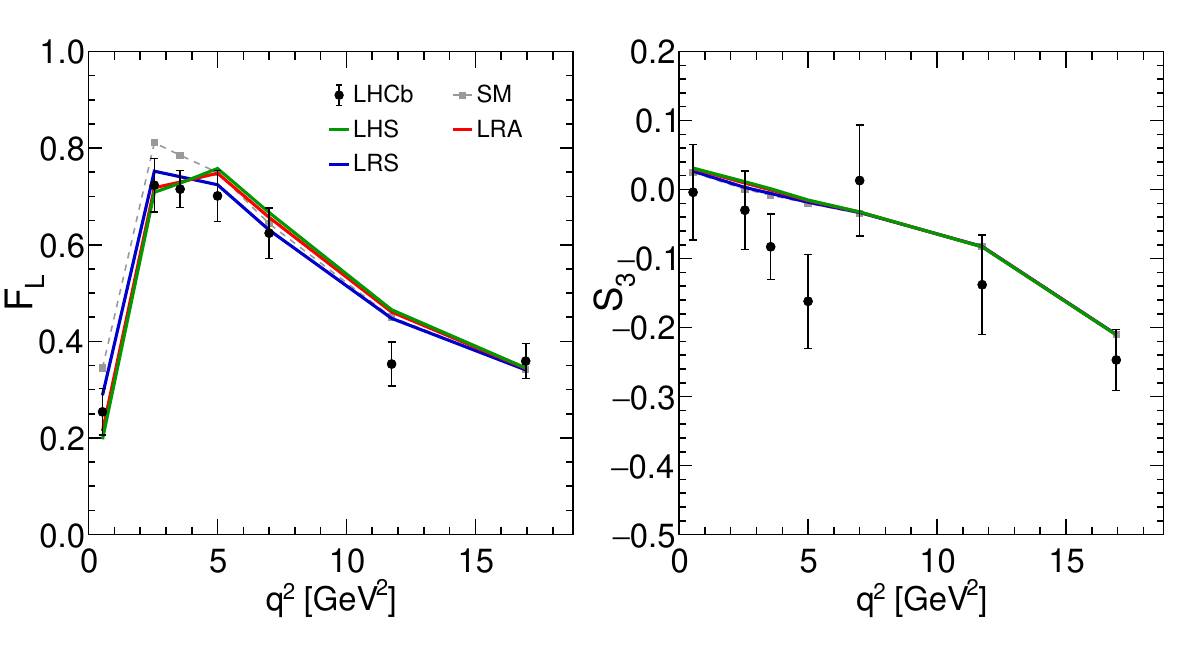}
    \caption{Longitudinal polarization fraction $F_L$ in $B_s \to \phi \mu^+ \mu^-$. and Angular observable $S_3$ in $B_s \to \phi \mu^+ \mu^-$. Consistency checks across angular observables in $B_s \to \phi \mu^+ \mu^-$.}
\label{fig:bphi_consistency}
\end{figure}

The longitudinal polarization fraction $F_L$ is well reproduced across all three scenarios, as shown in Fig.~\ref{fig:bphi_consistency}. In contrast, the observable $S_3$ exhibits a poorer level of agreement. While the model predictions remain close to the SM expectation, the experimental measurements show larger fluctuations. However, it is important to note that the theoretical predictions for $S_3$ (in $B \to \phi \mu^+ \mu^-$) carry significant uncertainties, arising from hadronic form factors and non factorizable QCD effects. A similar situation explains the poor fit in $S_9$ in $B \to K \mu^+ \mu^-$ 

As a result, the apparent tension in $S_3$ ($B \to \phi \mu^+ \mu^-$) and $S_9$ ($B \to K \mu^+ \mu^-$) should be interpreted with caution, as it may reflect limitations in theoretical precision rather than a clear indication of missing NP contributions.

\section{Extension to Neutrinos}
To ensure consistency with the electroweak gauge structure, we extend the $Z'$ interaction to include neutrinos. In the SM, the left handed muon and its associated neutrino form an $SU(2)_L$ doublet,
\[
\begin{pmatrix}
\nu_\mu \\
\mu
\end{pmatrix}_L,
\]
and any $U(1)'$ extension must respect this symmetry. As a result, the $Z'$ coupling to left handed muons necessarily implies a corresponding coupling to left handed neutrinos with equal strength $g_\nu^L = g_\mu^L$. Since neutrinos in the SM are purely left handed, no right handed coupling is introduced, and the interaction is entirely chiral.

\begin{equation}
\begin{aligned}
        \mathcal L_{NP} \supset &-\frac{1}{4} Z'_{\mu\nu}Z'^{\mu\nu} + \frac{1}{2}M_{Z'}^2Z'_{\mu}Z'^{\mu} + \\
        &\bar \chi(i \gamma^\mu \partial_\mu)\chi + g_\chi \bar \chi \gamma^\mu \chi Z'_\mu + \\
        &\bar s\gamma^\mu(g^L_{bs}P_L + g_{bs}^R P_R)bZ'_\mu  + \\
        & \bar \mu \gamma^\mu(g_\mu^LP_L + g_\mu^RP_R) \mu Z'_\mu  + \\
        & \bar\nu_\mu(g^L_{\nu_\mu}\gamma^\mu P_L)\nu_\mu
\end{aligned}
\end{equation}

This extension provides an important complementary probe of the chiral structure of the $Z'$ couplings. While observables such as $P'_5$ in $B \to K^* \mu^+ \mu^-$ primarily constrain the vector combination $(g_\mu^L + g_\mu^R)$, and $B_s \to \mu^+ \mu^-$ constrains the axial combination $(g_\mu^L - g_\mu^R)$, these measurements alone do not uniquely determine $g_\mu^L$ and $g_\mu^R$ individually.

In contrast, processes involving neutrinos, such as $B \to K \nu \bar{\nu}$, depend only on the left handed coupling,
\[
\text{BR}(B \to K \nu \bar{\nu}) \propto (g_{bs}^V)^2 (g_\nu^L)^2,
\]
and therefore directly constrain $g_\mu^L$ through the relation $g_\nu^L = g_\mu^L$.
\section{Quantifying Lepton Sector Chirality}
The global fit results indicate that the LRA configuration provides the best agreement with experimental data. This scenario allows independent variation of left- and right-handed muon couplings to the $Z'$, thereby introducing additional chiral freedom in the lepton sector. We therefore proceed to quantify the degree of asymmetry between the couplings $g_\mu^L$and $g_\mu^R$.

To this end, a profile likelihood analysis was performed in which the parameter of interest is the muon chiral factor $\theta \equiv f_\mu = \frac{g_\mu^R}{g_\mu^L}$

\begin{equation}
    \mathcal{L}(\theta, \eta, \phi) = \mathcal{L}(f_\mu: g_{\mu,0}, g_{bs,0} ; m_{Z'}, m_\chi, g_\chi)
\end{equation}
Nuisance parameters, including the overall muon coupling $g_{\mu,0}$ and the overall quark coupling $g_{bs,0}$, were profiled out. The remaining inputs were held fixed at $m_{Z'}=3\,\mathrm{TeV}$, $m_{\chi}=1\,\mathrm{GeV}$, $g_{\chi}=0.2$, and $g_{bs}^{R}/g_{bs}^{L}=0.087$. The fermion couplings are parametrized as
\begin{equation}
\begin{aligned}
g_\mu^L &= g_{\mu,0}, &
g_\mu^R &= f_\mu\, g_{\mu,0}, \\
g_{bs}^L &= g_{bs,0}, &
g_{bs}^R &= 0.087\, g_{bs,0}.
\end{aligned}
\end{equation}

The extension to neutrinos provides an important complementary probe of the chiral structure of the $Z'$ couplings. A key challenge in determining the chiral structure arises from degeneracies between vector and axial contributions. In particular, angular observables such as $P_5'$ predominantly probe the vector combination $g_\mu^L + g_\mu^R$, while the branching ratio $BR(B_s \to \mu^+ \mu^-)$ is sensitive to the axial combination $g_\mu^L - g_\mu^R$. To break this degeneracy, we incorporate additional constraints from neutrino modes. Under $SU(2)_L$ gauge invariance, the left handed muon and its associated neutrino form a doublet, implying $g_\nu^L = g_\mu^L$. This allows observables such as $BR(B \to K \nu \bar\nu)$, which depend only on left handed couplings, to directly constrain $g_\mu^L$ and thereby enable a combined determination of both chiral components.

\begin{figure}
 	  \centering
    \includegraphics[width=\linewidth]{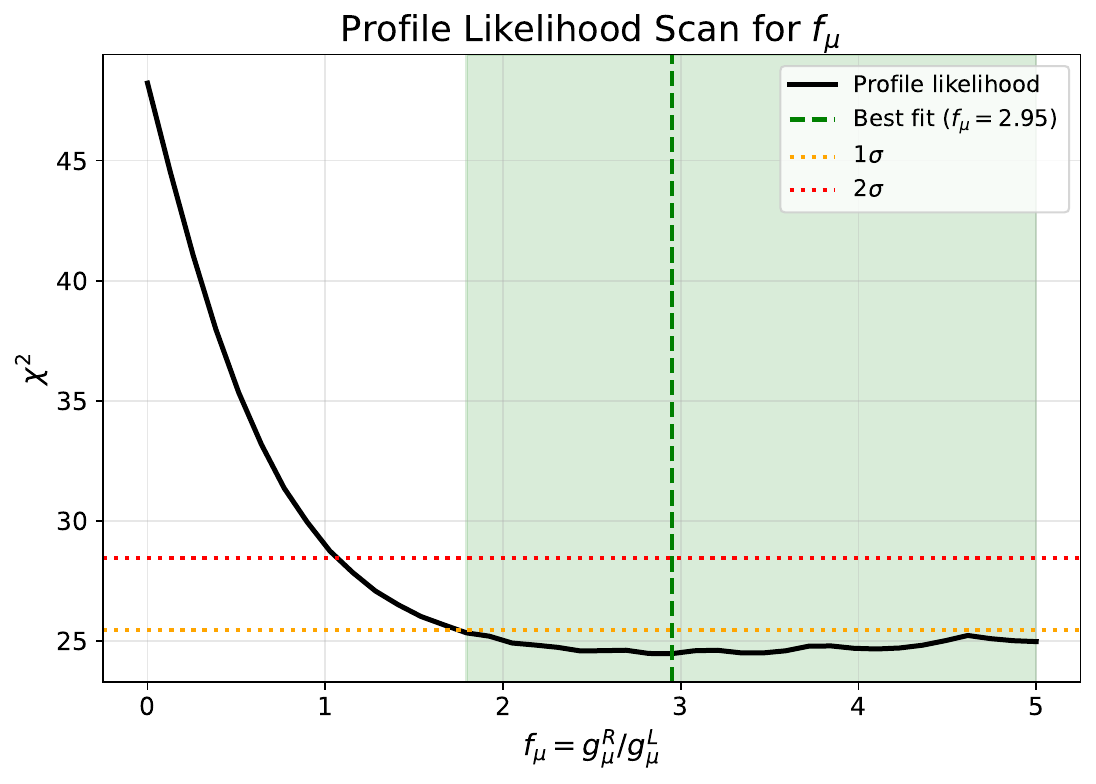}
     \caption{Profile likelihood scan for the muon chiral factor $f_\mu = g_\mu^R / g_\mu^L$, showing a best fit value near $f_\mu \approx 2.95$ with a well defined $1\sigma$ interval.}
    \label{fig:fmu-scan}
\end{figure}

\begin{figure}
 	  \centering
                \includegraphics[width=\linewidth]{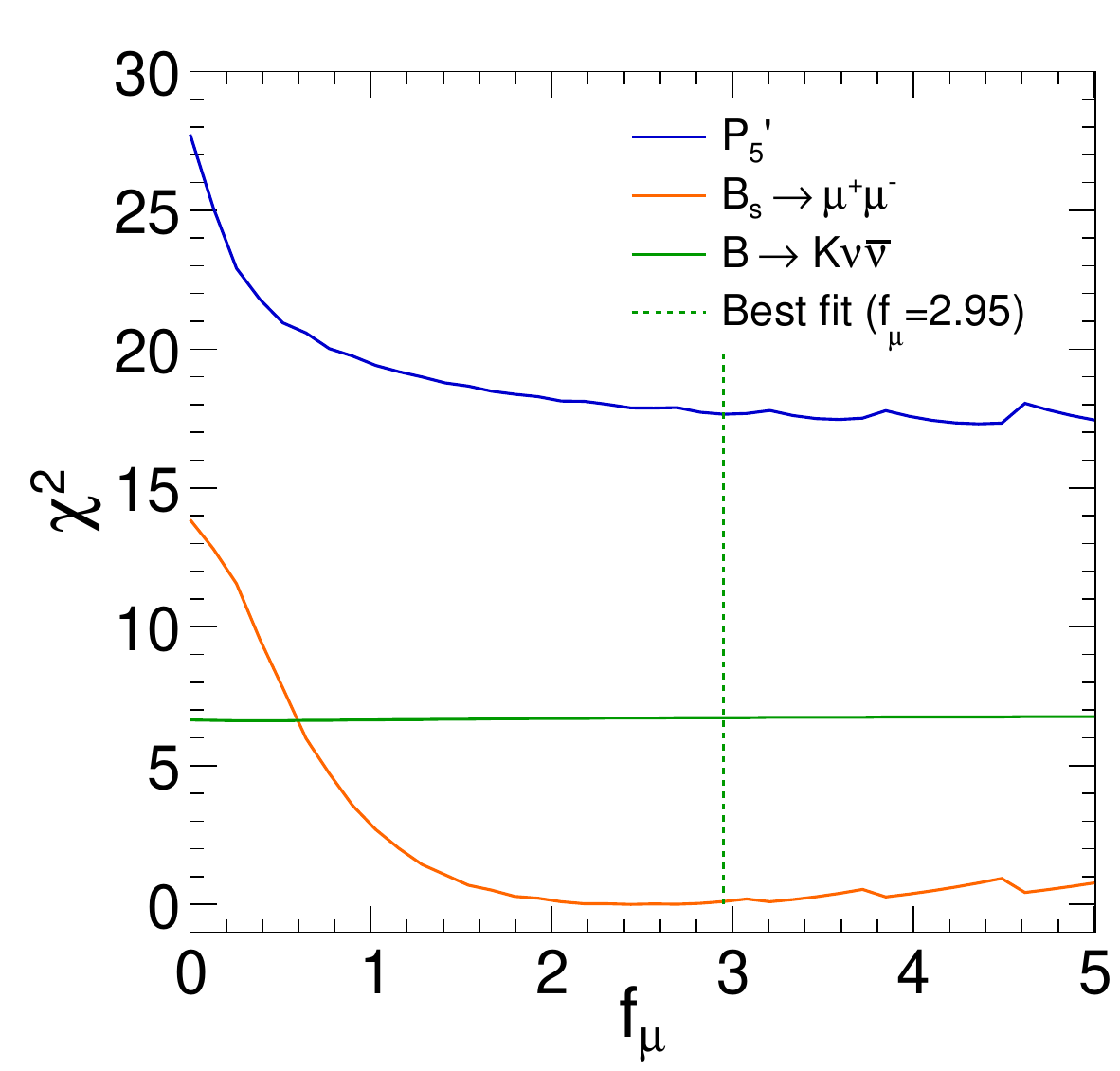}
                 \caption{Contributions of individual observables to the total $\chi^2$, illustrating how vector ($P'_5$), axial ($B_s \to \mu^+\mu^-$), and neutrino ($B \to K\nu\bar{\nu}$) channels jointly constrain the chiral structure.}
    \label{fig:triangulation}
\end{figure}

\begin{figure}
 	  \centering
         \includegraphics[width=\linewidth]{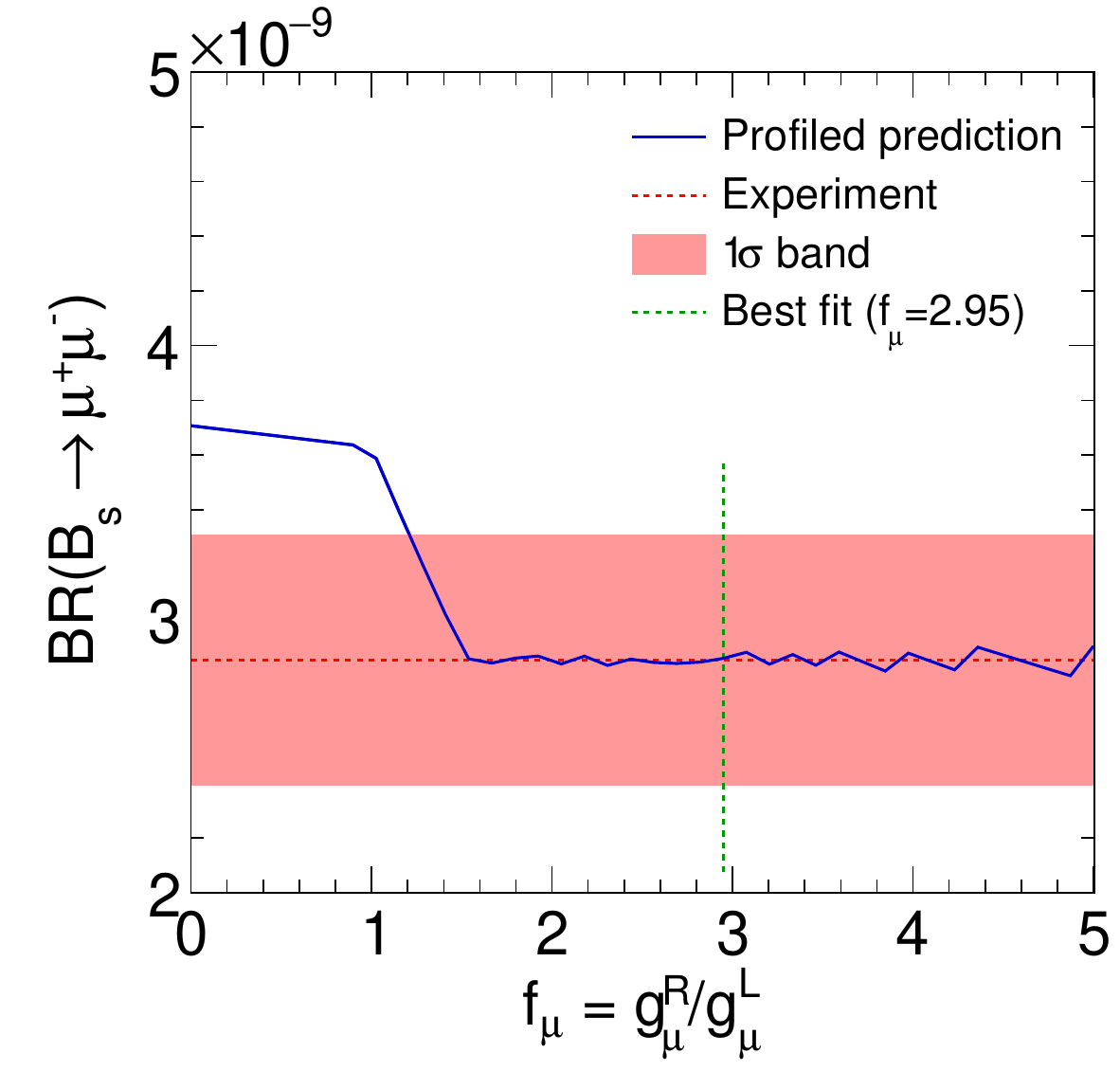}
        \caption{Dependence of $\mathrm{BR}(B_s \to \mu^+\mu^-)$ on $f_\mu$, showing consistency with experimental constraints at the preferred chiral asymmetry.}
    \label{fig:bsmumu-fmu}
\end{figure}

Using the three channel triangulation, we perform a profile likelihood scan over $f_\mu$. 
The profile likelihood is shown in Fig.~\ref{fig:fmu-scan}, the contribution of each channel to the total $\chi^2$ is shown in Fig.~\ref{fig:triangulation}, and the dependence of $\mathrm{BR}(B_s\to\mu^+\mu^-)$ on $f_\mu$ is shown in Fig.~\ref{fig:bsmumu-fmu}.
The resulting best fit value is found to be:
\begin{equation*}
f_\mu^{\text{best}} = 2.95^{+1.87}_{-1.16}
\end{equation*}
corresponding to a $1 \sigma$ interval defined by $\Delta \chi^2 \leq 1$ for a single parameter. This result indicates a clear preference for a hierarchy in which the right handed coupling is enhanced relative to the left handed one.

This asymmetry plays a crucial role in achieving a consistent fit across observables. In particular, it enables the model to simultaneously accommodate the vector dominated anomaly in $P_5'$ (primarily sensitive to $C_9$) and the axial sensitive constraint from $BR(B_s \to \mu^+ \mu^-$) (dominated by $C_{10}$). The introduction of independent chiral degrees of freedom therefore provides the flexibility required to reconcile tensions across multiple decay channels within a unified framework.

\section{Summary }
\label{Summary}
The role of momentum dependent effects in $Z'$ portal models addressing anomalies in rare \(B\) meson decays was investigated. By incorporating vacuum polarization corrections arising from a dark fermion, a modified $Z'$ propagator was constructed and its impact on the Wilson coefficient $C_9$ and related observables was studied. Different chiral realizations of the $Z'$ boson and different coupling strengths to right handed and left handed currents were explored. All three models led to a substantial improvement in the fit to the angular observable $P_5'$, with reductions in the tension in the central energy bin of $[1.1,6.0]~\mathrm{GeV}^2$, where the anomaly is most prominent, quantified as approximately $92\%$ (LHS), $96\%$ (LRA), and $76\%$ (LRS).

The LRA scenario emerges as the most consistent overall framework, as it introduced sufficient chiral freedom to independently adjust vector and axial contributions. This allowed for the simultaneous accommodation of both angular and branching ratio observables. A profile likelihood analysis identified a preferred hierarchy in the lepton sector, $g_\mu^R \approx 2.95g_\mu^L$. The preferred hierarchy indicated that purely left handed scenarios were insufficient to simultaneously accommodate both vector sensitive observables, such as $P_5'$, and axial constraints from decays such as $B_s \to \mu^+\mu^-$.

The vacuum polarization contribution introduced a non trivial $q^2$ dependence into the effective interaction. Although subleading in magnitude, the correction generated a threshold feature near $q^2 \sim 4m_\chi^2$, which overlapped with the kinematic region where deviations were observed. The observed threshold behavior suggested that even small loop induced effects can modify the structure of the effective theory in a way that is not captured by constant shifts in Wilson coefficients.

From a theoretical perspective, the results demonstrated that radiative corrections in $Z'$ portal models are not optional additions but arise naturally from the underlying interactions. Their inclusion provided a more complete description of the effective propagator and introduced energy dependent structure that may be relevant for precision flavor observables.

Overall, this study supported the view that both chiral structure and momentum dependent corrections should be taken into account in phenomenological analyses of flavor anomalies, particularly in frameworks that couple the Standard Model to a dark sector.

The present framework demonstrated how momentum dependent effects from dark sector loops can be incorporated consistently into $Z'$ portal descriptions of rare \(B\) meson decays. While the dominant contribution to the observables remained the tree level mediator exchange, the vacuum polarization correction introduced additional structure into the effective interaction, particularly near threshold regions.

The analysis was formulated specifically for $\Delta B = 1$ flavor changing neutral current processes relevant to rare semileptonic decays. Consequently, $\Delta B = 2$ observables such as $B_s-\bar{B}_s$ mixing were not addressed directly, as their treatment would require an extended operator basis and a broader global analysis.

An important outcome of the analysis was the emergence of a preferred left right asymmetric lepton sector configuration. The identified configuration suggested that allowing independent chiral couplings may provide additional flexibility in simultaneously accommodating vector sensitive angular observables and branching ratio constraints within a unified framework. The analysis was performed using a discrete parameter space exploration. A natural next step would therefore be the implementation of a continuous likelihood optimization or a full global fit, which could more precisely constrain the allowed parameter regions and correlations between couplings.

Future high precision measurements with finer $q^2$ resolution at LHCb Run 3 and Belle II may provide increased sensitivity to localized momentum dependent features such as the threshold behavior identified in the present study.

\begin{acknowledgement}
S. Gupta gratefully acknowledges Ms. Monica Gupta and Bhuvan Gupta for their valuable contributions and their support throughout this work.
S. Dogra acknowledges the financial support provided by the National Research Foundation of Korea (NRF) under project numbers. RS-2008-NR007227 and RS-2025-000561200.
\end{acknowledgement}

\bibliographystyle{unsrt}
\bibliography{references}
\end{document}